\documentclass{aa}  

\usepackage{graphicx}
\usepackage{txfonts}
\usepackage{subcaption}         

\usepackage{amsmath}
\usepackage{xcolor}
\usepackage{empheq}
\usepackage{float}
\usepackage{makecell}
\usepackage{booktabs}
\usepackage{array}

\usepackage[pdfencoding=auto]{hyperref}
\hypersetup{
    colorlinks=true,
    linkcolor=blue,
    filecolor=magenta,      
    urlcolor=blue,
    citecolor=blue
}
\begin{document}

\newcommand{\abz}{\alpha_{\mathrm B0} }
\newcommand{\amz}{\alpha_{\mathrm M0} }
\newcommand{\ab}{\alpha_{\mathrm B} }
\newcommand{\am}{\alpha_{\mathrm M} }
\newcommand{\nlt}[2]{\langle #1, #2 \rangle}

   \title{\texttt{PySCo-EFT} and \texttt{ECOSMOG-EFT}: a tandem of $N$-body simulation codes for the Effective Field Theory of Dark Energy}
   \titlerunning{EFTofDE Simulations with \texttt{PySCo-EFT} and \texttt{ECOSMOG-EFT} }



   \author{H. Ganjoo\inst{1} \corrauth{Himanish.Ganjoo@obspm.fr}, Y. Rasera\inst{1,2,3}\mailadd{Yann.Rasera@obspm.fr}
   , E. Bellini \inst{4},
        M-A. Breton \inst{5}, F. Castillo \inst{1}, S. Codis \inst{5}, S. Colombi \inst{6}, P.-S. Corasaniti \inst{1,6}, G. Cusin \inst{6}, Y. Dubois \inst{6}, S. de la Torre \inst{7}, E. Jullo \inst{7}, G. Lavaux \inst{6}, A. Le Brun \inst{1}, S. Peirani \inst{6,8,14}, S. Pires \inst{5}, V. Reverdy \inst{9}, I. Sáez-Casares \inst{10,15}, S. Saga \inst{11,12}, A. Taruya \inst{13,14}, W. Zhang \inst{6}}
    \authorrunning{H. Ganjoo et al.}
   \institute{LUX, Observatoire de Paris, Université PSL, Sorbonne Université, CNRS, 92190 Meudon, France
            \and Université Paris Cité, F-75006 Paris
            \and Institut Universitaire de France
            \and \textit{Center for Astrophysics and Cosmology}, University of Nova Gorica, Nova Gorica, Slovenia
            \and Universit\'e Paris-Saclay, Universit\'e Paris Cit\'e, CEA, CNRS, AIM, 91191 Gif-sur-Yvette, France
            \and Institut d'Astrophysique de Paris, CNRS and Sorbonne Universit\'e, UMR 7095, 98 bis Boulevard Arago, F-75014 Paris, France
            \and Aix-Marseille Université, CNRS, CNES, LAM, Marseille, France
            \and ILANCE, CNRS – University of Tokyo International Research Laboratory, Kashiwa, Chiba 277-8582, Japan
            \and Laboratoire d’Annecy de Physique des Particules (LAPP), CNRS-IN2P3, 9 Chemin de Bellevue - BP 110, 74941 Annecy Cedex, France
            \and Dipartimento di Fisica `Aldo Pontremoli', Università degli Studi di Milano, Via Celoria 16, I-20133 Milan, Italy
            \and Institute for Advanced Research, Nagoya University, Furo-cho, Chikusa-ku, Nagoya 464-8601, Japan
            \and Kobayashi-Maskawa Institute for the Origin of Particles and the Universe, Nagoya University, Furo-cho, Chikusa-ku, Nagoya, 464-8602, Japan
            \and Center for Gravitational Physics and Quantum Information, Yukawa Institute for Theoretical Physics, Kyoto University, Kyoto 606-8502, Japan
            \and Kavli IPMU (WPI), UTIAS, The University of Tokyo, Kashiwa, Chiba 277-8583, Japan
            \and INFN, Sezione di Milano, Via Celoria 16, I-20133 Milan, Italy
            }
   \date{Received XXX}

 
  \abstract
   {
   Modified gravity theories constitute viable alternatives to the standard cosmological model for explaining the observed late-time accelerated expansion of the Universe. The Effective Field Theory of Dark Energy (EFTofDE) is an efficient framework to describe a wide range of such theories with a limited number of parameters. To robustly constrain them, high-resolution cosmological $N$-body simulations are required to obtain accurate predictions for the matter distribution on non-linear scales, which can then be quantitatively compared with clustering and weak gravitational lensing measurements from forthcoming galaxy surveys.
   We introduce two new $N$-body simulation codes for cosmologies governed by EFTofDE: \texttt{PySCo-EFT}, a fast Python-based particle mesh code, and \texttt{ECOSMOG-EFT}, an accurate \texttt{RAMSES}-based code with adaptive mesh refinement.  
   We consider Horndeski models with a luminal gravitational wave speed. We use iterative solvers and multigrid schemes to solve for the additional scalar field equation in both codes, including the non-linear Vainshtein screening mechanism. The standard gravitational force is augmented by a fifth force from the scalar field.
   We present several validation and convergence tests of the codes. We obtain a sub-0.5 percent agreement with linear theory on large scales and a similar agreement between the two codes in the non-linear regime although they implement different solvers. The dominant numerical effects on the matter-power-spectrum boost are finite mass resolution, finite-volume effects, refinement threshold, and starting redshift, but they are limited to below 2\% at the largest wavenumbers $k=10\,h^{-1}\mathrm{Mpc}$ for the range of tested values. We finally investigate the impact of the EFTofDE parameters on the matter-power-spectrum ratios between EFTofDE and $\Lambda$CDM scenarios. Depending on the values of the EFTofDE parameters, the screening can play a negligible or dominant role compared to the often-used linearised field equations.
   Our codes provide tools for generating fast and accurate predictions of the impact of the EFTofDE on the clustering of matter, incorporating the non-linear Vainshtein screening mechanism.
   }
   

\maketitle
\nolinenumbers
\section{Introduction}

Despite the success of the standard $\Lambda$CDM model of cosmology at explaining a wide array of observations with cold dark matter (CDM) and a cosmological constant ($\Lambda$) responsible for the accelerated expansion of the Universe at late times, the exact nature of the so-called ``dark energy'' remains elusive. Moreover, recent results from large-scale structure surveys like the Dark Energy Spectroscopic Instrument \citep[DESI;][]{DESI:2025zgx} and the Dark Energy Survey \citep[DES;][]{DES:2025bxy} indicate a preference towards a time-evolving dark energy component over a cosmological constant. This gap in our understanding has led to a host of theories as alternatives to $\Lambda$, including modified gravity (MG) models. 

Observations of the large-scale structure (LSS) in the next decade will serve as a laboratory for investigating the nature of gravity on cosmological scales, providing an excellent method to constrain MG theories. Stage-IV LSS surveys like \textit{Euclid} \citep{Euclid1}, DESI \citep{DESI1}, and the Large Synoptic Survey Telescope \citep{LSST2012} will yield data with unprecedented volume and depth, enabling the generation of more precise statistics than ever before. Constraining MG theories with such precise data requires accurate predictions, which account for the impact of MG on the LSS from linear to non-linear scales.

The Effective Field Theory of Dark Energy (EFTofDE) provides an efficient framework to unify the description of a wide range of MG scenarios using a minimal set of parameters \citep{Gubitosi13_EFTofDE,Bloomfield:2012ff,Bellini:2014fua,Frusciante20_EFTofDE_review}. This has enabled the development of Einstein--Boltzmann solvers for the EFTofDE, such as \texttt{hi\_class} \citep{Zumalacarregui:2016pph,Bellini:hiclass} and EFTCAMB \citep{Hu14_EFTCAMB}, both of which have been validated for the precision needed by current and forthcoming surveys \citep{Bellini:2017avd}. While useful for cosmic microwave background (CMB) analysis, these solvers are limited to the linear regime of structure formation. The EFTofDE has been further extended to account for non-linearities in the field equations \citep[e.g.][]{Bellini15_bispectrum,cusin,Frusciante17_nonlinearEFTofDE}. The linear predictions have also been extended to the quasi-linear regime or non-linear regime using several approaches either based on perturbation theory \citep{Bellini15_bispectrum,Yamauchi17_bispectrum,Cusin18_EFTofLSS}
or the halo model \citep{Bose23_halomodel_EFTofDE,deBoe_halomodel_EFTofDE}. An alternative path for more accuracy is to use $N$-body solvers. Several particle-mesh codes have been developed for parameterised MG \citep{Hassani:2019lmy,Hassani:2020rxd,Ruan:2021wup,Wright:2022krq,Gupta:2024seu,Brando,Nouri-Zonoz:2025cul} to explore the linear and quasi-linear regimes of these models. On the other hand, the cubic and quartic Galileon models, implemented via similar equations to the ones of the EFTofDE, have previously been implemented in adaptive mesh $N$-body simulation codes \citep{Barreira:2013eea, Becker:2020azq}.

In this work, we present two $N$-body simulation codes designed to run simulations of cosmic structure formation for MG scenarios described by the EFTofDE, including the Vainshtein screening mechanism that recovers standard gravity on small scales in order to satisfy constraints on gravity from local tests \citep{Will2014}.
The equation for the additional scalar field that implements Vainshtein screening is non-linear and contains products of the derivatives of the field. Fourier-based solvers and tree methods cannot be used to solve this equation at very small scales. 
This computational difficulty led to our choice of grid-based codes employing multigrid solvers for this project. We added the EFTofDE formalism to the fast and versatile Python-based particle mesh code \texttt{PySCo} \citep{pysco} for quick preliminary simulations, and to \texttt{ECOSMOG-CVG} \citep{ecosmog,Becker:2020azq}, a \texttt{RAMSES}-based \citep{teyssier02} code, for high-resolution simulations with adaptive mesh refinement (AMR). We chose to implement a subset of the Horndeski class of MG models, in which an additional scalar field couples to the spacetime metric \citep{Horndeski:1974wa,Deffayet:2011gz}. Moreover, we parameterised the Horndeski Lagrangian in the $\alpha$-basis \citep{Bellini:2014fua}, in which a few time-dependent functions can codify the EFTofDE. The resulting $N$-body codes are called \texttt{PySCo-EFT}\footnote{PySCo-EFT: 
\\
\url{https://github.com/hganjoo/pyscoeft/} \\
\url{https://doi.org/10.5281/zenodo.19496403}} and \texttt{ECOSMOG-EFT}\footnote{ECOSMOG-EFT: 
\\
\url{https://github.com/hganjoo/ecosmogeft/} \\
\url{https://doi.org/10.5281/zenodo.19496317}}. 

To the best of our knowledge, our codes provide the first $N$-body simulation tools to explore the space of EFTofDE models in the $\alpha$-basis including the non-linear screening terms. Up to this date, there exist no Python-based or AMR implementations of these models for $N$-body simulations. This enables the generation of precise predictions on linear and non-linear scales to directly constrain the $\alpha$ functions, which quantify the physically relevant properties of Horndeski theories. Additionally, the scalar field modifies both the gravitational potentials $\Psi$ and $\Phi$. While structure formation is only affected by $\Psi$, lensing signatures depend on $\Psi + \Phi$. Implementing this EFTofDE model in an AMR-based code enables us to accurately compute both these fields in highly dense environments, enhancing the ability to generate lensing signatures from structures on galaxy scales.

This paper is organised as follows. In Sec.~\ref{sec:theory}, we introduce the EFTofDE formalism and derive the equations we have implemented. We include both the linearised and the full non-linear equations and derive the linear growth factor for the EFTofDE cosmologies studied in this work. In Sec.~\ref{sec:methods}, we describe the numerical methods used and the modifications to both \texttt{PySCo} and \texttt{ECOSMOG}. We also enumerate the setups for the tests we conducted to validate the codes and to conduct a physical study of the impact of EFTofDE parameters on the matter power spectrum. In Sec.~\ref{sec:results}, we present the results of the aforementioned tests and simulations. We present our conclusions and perspectives for future work in Sec.~\ref{sec:conclusions}. Additionally, the methods and results for a suite of convergence tests are presented in Appendix~\ref{sec:convtests_methods} and Appendix~\ref{sec:convtests_results} respectively.

\section{Theory}
\label{sec:theory}

\subsection{The effective field theory of dark energy}

Beyond a simple cosmological constant, the observed late-time accelerated expansion of the Universe can be explained either by introducing a new dynamical component with negative pressure, commonly referred to as dark energy (DE), or by modifying the laws of gravity on cosmological scales (modified gravity, MG). A powerful and unifying way to describe both possibilities is provided by the Effective Field Theory of Dark Energy (EFTofDE), which offers a model-independent framework to parameterise deviations from general relativity at the level of cosmological perturbations while retaining some level of control over theoretical consistency \citep{Gubitosi13_EFTofDE,Bloomfield:2012ff,Bellini:2014fua,Frusciante20_EFTofDE_review}.

The EFTofDE is constructed by writing the most general action for perturbations around a Friedmann--Lema\^{i}tre--Robertson--Walker (FLRW) background that is invariant under time-dependent spatial diffeomorphisms, assuming the presence of a single additional scalar degree of freedom. This approach encompasses a wide class of scalar-tensor theories, including Horndeski gravity and several of its well-studied subclasses, while allowing one to remain agnostic about the underlying covariant Lagrangian.

At the background level, the EFTofDE framework reproduces a generic cosmological expansion history, which can be chosen to match that of the standard cosmological model ($\Lambda$CDM) or generalised to include a dynamical dark energy equation of state with arbitrary time dependence. Deviations from general relativity arise in the evolution of perturbations and are encoded in a minimal set of time-dependent functions that govern the dynamics of both scalar and tensor modes. This separation between background evolution and perturbative effects makes the EFT-based approach particularly well suited for testing general deviations from $\Lambda$CDM+GR with the CMB and LSS observables.

A convenient and widely adopted parametrization of the EFTofDE action is given by the so-called $\alpha$-basis, in which the effects of MG and DE are described by one function of time describing the background dynamics plus four dimensionless functions fixing the evolution of linear perturbations. While this is strictly valid for the Horndeski class of models, which are the focus of this work, generalizations to beyond-Horndeski \citep{Zumalacarregui:2013pma,Gleyzes:2014qga,Gleyzes:2014rba} and DHOST \citep{Langlois:2015cwa,Langlois:2017mxy,Langlois:2018dxi} theories have been studied, at the cost of introducing a few additional functions of time. These functions control physically transparent properties of the theory (see \citealp{Bellini:2014fua} for a detailed description of these functions), such as effective Planck mass running rate ($\alpha_{\rm M}$), the kinetic mixing between the scalar field and the metric, or the braiding ($\alpha_{\rm B}$), the propagation of gravitational waves ($\alpha_{\rm T}$), and the kinetic energy of scalar perturbations ($\alpha_{\rm K}$). This formulation has proven especially useful for connecting theoretical models to cosmological observables and for implementing MG consistently in Einstein--Boltzmann solvers and $N$-body simulations \citep{Zumalacarregui:2016pph,Bellini:hiclass,Hassani:2019lmy,Nouri-Zonoz:2025cul}.

On non-linear scales, additional operators appear in the EFT action \citep{Bellini15_bispectrum,cusin}, giving rise to screening mechanisms (controlled by $\alpha_{\rm V1}$ and $\alpha_{\rm V2}$ functions, following the definition given in \citealp{cusin}) that suppress deviations from general relativity in high-density or small-scale environments. These effects are essential for ensuring compatibility with local gravity tests and play a crucial role in determining the impact of MG on structure formation. In this work, we focus on a minimal yet non-trivial subset of the EFTofDE that captures the leading non-linear interactions relevant for cosmological $N$-body simulations, while remaining consistent with current observational constraints.

Below, we introduce the explicit form of the EFT action adopted in this work and derive the corresponding equations of motion under the quasi-static approximation.

\subsection{The non-linear effective field theory of dark energy action}

This section is a quick introduction to the non-linear EFTofDE
action. We shall adopt the notations proposed in \citet{cusin}, to which we refer for additional details. We consider non-relativistic gravitational fields and velocities and we concentrate on scales much shorter than the Hubble radius, where relativistic effects due to the expansion of the Universe can be neglected. For gravitational and field fluctuations
below the scalar field sound horizon, we can assume the quasi-static approximation and time
derivatives can be taken to be much smaller than spatial derivatives. 

In \citet{cusin}, it has been shown that in the quasi-static limit, there are only two additional independent operators to the quadratic action: one appearing at cubic order and the second at quartic order, in such a way that in the quasi-static limit the Horndeski Lagrangian can be written in terms of six independent operators only. The EFTofDE Lagrangian can be rewritten in the $\alpha$-basis to be represented by
a set of time-dependent functions $\{\alpha_{\rm M}, \alpha_{\rm B}, \alpha_{\rm T}, \alpha_{\rm V1}, \alpha_{\rm V2}, \alpha_{\rm V3}  \}$ and the time-varying Planck mass $M(t)$. 
The action can be expressed as a sum of terms containing increasing orders of the products of the different fields,
\begin{equation}
    S = S_{\rm m}+ S^{(2)}_{\rm g} + S^{(3)}_{\rm g} + ... ,
\end{equation} where $S_{\rm m}$ represents the coupling of matter to the metric and the $S^{(i)}_{\rm g}$ terms contain the field perturbations (at order $i$). To derive the perturbation equations from this Lagrangian, we will work in Newtonian gauge with the perturbed FLRW metric in natural units ($c=1$) considering only scalar fluctuations, \begin{equation}
    {\rm d}s^2 = -(1 + 2\Psi)\,{\rm d}t^2 + a^2(t)\,(1 - 2\Phi)\,\delta_{ij}\,{\rm d}x^i\,{\rm d}x^j,
\end{equation} where $\Psi$ and $\Phi$ are the two Bardeen potentials\footnote{Note that the $\Phi$ and $\Psi$ we use are switched compared to the notation in \citet{Cusin18_EFTofLSS}.}, $a$ is the expansion factor, and $\delta_{ij}$ is the Kronecker delta.
Considering scales much smaller than the Hubble radius and using the quasi-static approximation in which time derivatives are much smaller than spatial derivatives, the various action terms can be expressed up to third order as \begin{align}
S^{(2)}_{\rm g} &= \int {\rm d}^4x \, a M^2 \, A_{ab} \, \nabla_i \varphi_a \, \nabla_i \varphi_b , \\ 
S^{(3)}_{\rm g} &= \int {\rm d}^4x \, a M^2 \, B_{abc} \,
  \epsilon^{ikm}\epsilon^{jlm}\,
  \nabla_i \varphi_a \, \nabla_j \varphi_b \, \nabla_k \nabla_l \varphi_c , \\
S_{\rm m} &= -\int {\rm d}^4x \, a^3 \, \bar\rho_{\rm m} \, \Psi \, \delta ,
\end{align} 
where $\varphi_a \equiv \{\Psi,\Phi, \chi\}$, $\chi$ is the scalar field representing the additional degree of freedom, $M$ is the time-dependent Planck mass,  $\bar{\rho}_{\rm m}$ is the background matter density in the Universe, $\delta$ is the matter overdensity, and $\epsilon^{ijk}$ is the Levi-Civita symbol. Each term $S^{(i)}_{\rm g}$ carries $i$ copies of the fields. The matrices $A_{ab}$ and $B_{abc}$ consist of dimensionless time-dependent elements that enable the expression of the Lagrangian terms as sums over the products of different fields. We have  
\begin{equation} A_{ab} =
\begin{pmatrix}
0 & 1 & -\alpha_{\rm B} \\
1 & -1 - \alpha_{\rm T} & \alpha_{\rm M} - \alpha_{\rm T} \\
-\alpha_{\rm B} & \alpha_{\rm M} - \alpha_{\rm T} & -\mathcal{C}_2
\end{pmatrix} ,
\end{equation} in which \begin{align}
  \mathcal{C}_2 = \alpha_{\rm T} - \alpha_{\rm M}
      + \alpha_{\rm B}(1 + \alpha_{\rm M})
      + (1 + \alpha_{\rm B})\frac{\dot{H}}{H^2}
      + \frac{\dot{\alpha}_{\rm B}}{H}
      + \frac{\bar{\rho}_{\rm m}}{2 H^2 M^2},
\end{align} where overdots denote time derivatives and $H$ is the Hubble parameter. The elements of $B_{abc}$ are similarly given by linear combinations of the various $\alpha$ functions and their time derivatives (note that the coefficients $\alpha_{\rm V1}$ and  $\alpha_{\rm V2}$ enter the expression of $B_{abc}$). 

The simultaneous detection of GW170817 and its associated $\gamma$-ray burst GRB170817A \citep{Ligo_gw170817} has strongly constrained the speed of gravitational waves to be the same as the speed of light for any cosmological analysis \citep{Baker:2017hug,Creminelli:2017sry,Ezquiaga:2017ekz,Sakstein:2017xjx}, which motivates our choice of $\alpha_{\rm T} = 0$ for this work. As shown in \citet{cusin}, this choice imposes that the parameters that control higher-order field perturbation terms in the EFTofDE action are also $\alpha_{\rm V1}= \alpha_{\rm V2}=\alpha_{\rm V3}=0$. 

Considering the Lagrangian up to third order in the fields, activates the lowest order screening terms in the equation for the scalar field. While this choice may seem restrictive, it captures the contribution of all the leading terms on sub-horizon scales. Indeed, inspecting the Horndeski Lagrangian (which is the fundamental model on which we base our analysis), it is possible to notice that only a finite number of space derivatives can appear in the equation of motion. This number is given by the structure of the Lagrangian, and for luminal Horndeski theories ($\alpha_{\rm T}=0$) this implies that leading order contributions are fully contained up to second-order perturbation theory (third-order Lagrangian). To express this concept in a quantitative way, the sub-horizon and quasi-static approximations imply that spatial derivatives are much larger than time derivatives and the Hubble scale, i.e.~$\nabla^2\Psi\gg \ddot{\Psi} \sim H^2 \Psi$ for any perturbation. Additionally, perturbations are constructed to be small, i.e.~$\Psi\ll 1$, so that, e.g., $\nabla^2\Psi\gg\partial_i\Psi\partial^i\Psi$. Then, the exercise is merely to count how many field perturbations can have each term of the Lagrangian and the corresponding number of space derivatives (remember that space derivatives act only at the perturbation level, the background is just time-dependent due to homogeneity and isotropy). 

This cubic screening model leaves two non-zero $\alpha$ parameters: $\ab$ and $\am$, such that $B_{333} = \mathcal{C}_4 \equiv -4\ab + 2\am$ is the only non-zero element in $B_{abc}$. Additionally, this choice yields three decoupled equations for the three fields, making the system amenable to a simple numerical method for obtaining solutions.

\subsection{Equations of motion}

Restricting the action to third order in field perturbations, the non-linear equations of motion for the three fields at second order can be obtained by varying the action \begin{align} \label{eq:action} 0 &= \frac{\delta (S^{(2)}_{\rm g} + S^{(3)}_{\rm g} + S_{\rm m})}{\delta \varphi_d} \\
  &= A_{da}\,\nabla^2 \varphi_a
  - \delta_{d1}\,\frac{\bar{\rho}_{\rm m} a^2}{2 M^2}\,\delta
  - \frac{B_{dab}}{4 H^2 a^2}\,\epsilon^{ikm}\epsilon^{jlm}\,
    \nabla_i \nabla_j \varphi_a\, \nabla_k \nabla_l \varphi_b,
\end{align} 
where $\bar{\rho}_{\rm m}$ is the background matter density at scale factor $a$. 

Equation~(\ref{eq:action}) yields three equations for the three fields, given by
\begin{subequations}\label{eq:field_eqs}
\begin{align}
\nabla^2 \Phi - \alpha_{\rm B} \nabla^2 \chi - \frac{\bar{\rho}_{\rm m} a^2}{2 M^2} \delta &= 0,
\label{eq:psi}\\[2pt]
\nabla^2 \Psi - \nabla^2 \Phi + \alpha_{\rm M} \nabla^2 \chi &= 0, 
\label{eq:phi}\\[2pt]
-\alpha_{\rm B} \nabla^2 \Psi
+ \alpha_{\rm M} \nabla^2 \Phi
- \mathcal{C}_2 \nabla^2 \chi
& \nonumber\\
\quad
- \frac{\mathcal{C}_4}{4 H^2 a^2}
\Big[
(\nabla^2 \chi)^2
- \nabla_i \nabla_j \chi\, \nabla^i \nabla^j \chi
\Big] &= 0,
\label{eq:chi_prelim}
\end{align}
\end{subequations} where the non-linear term $\epsilon^{ikm} \epsilon^{jlm} \nabla_i \nabla_j \chi \nabla_k \nabla_l \chi$ has been expanded using $\varepsilon^{ikm} \varepsilon^{jlm} = \delta^{ij} \delta^{kl} - \delta^{il} \delta^{kj}$. This non-linear system can be rearranged to obtain an equation of motion for $\chi$ that depends only on $\chi$ and its derivatives along with the overdensity $\delta$, 
\begin{equation} \label{eq:chi} \boxed{\begin{aligned}
    (2 \ab \am - \ab^2 - \mathcal{C}_2) \nabla^2 \chi + (\am - \ab) \frac{\bar{\rho}_{\rm m} a^2}{2 M^2} \delta \\ 
    -\frac{\mathcal{C}_4}{4H^2 a^2}\left[
\left(\nabla^2 \chi\right)^2 
- \nabla_i \nabla_j \chi \, \nabla^i \nabla^j \chi
\right] = 0.
\end{aligned}}
\end{equation} 
The third term on the left-hand side of Eq.~(\ref{eq:chi}) implements the Vainshtein screening mechanism, by which the effect of MG is reduced at small scales. The $\mathcal{C}_4$ term controls the strength of the screening.  Additionally, the equations for $\Psi$ and $\Phi$ are given by
\begin{subequations} \label{eqs:phipsi}
    \begin{empheq}[box=\fbox]{align}
  \nabla^2 \Psi &=  \frac{\bar{\rho}_{\rm m} a^2}{2 M^2} \delta + (\ab - \am)\nabla^2 \chi, \\
  \nabla^2 \Phi &=  \frac{\bar{\rho}_{\rm m} a^2}{2 M^2} \delta + \ab \nabla^2 \chi.
\end{empheq}
\end{subequations}

Equations~(\ref{eqs:phipsi}) also demonstrate how the scalar field $\chi$ modifies the evolution of the potentials $\Psi$ and $\Phi$: the source terms get additional contributions from $\nabla^2 \chi$ with prefactors that are functions of the $\alpha_{\rm M}$ and $\alpha_{\rm B}$. Thus, the scalar field creates a slip between $\Psi$ and $\Phi$.

\subsection{Time dependence of the $\alpha$ functions}

For this work, we chose to follow the same parameterisation as in \citet{cusin} so that \begin{equation} \label{eq:alpha-a}
    \alpha_{I} (a) = \alpha_{I0} \frac{1 - \Omega_{\rm m}(a)}{1 - \Omega_{\rm m0}} \, , \, \Omega_{\rm m}(a) = \frac{\Omega_{\rm m0}}{\Omega_{\rm m0} + (1 - \Omega_{\rm m0})(a/a_0)^3},
\end{equation} 
where $I=M,B$ and $\Omega_{\rm m0}$ is the present-day matter density normalised by the critical density.
This functional form makes the $\alpha$ values scale with the proportion of DE in the Universe, such that the impact of the scalar field is relevant during DE domination at late times.

\begin{figure}[h!]
   \centering
   \includegraphics[width=\hsize]{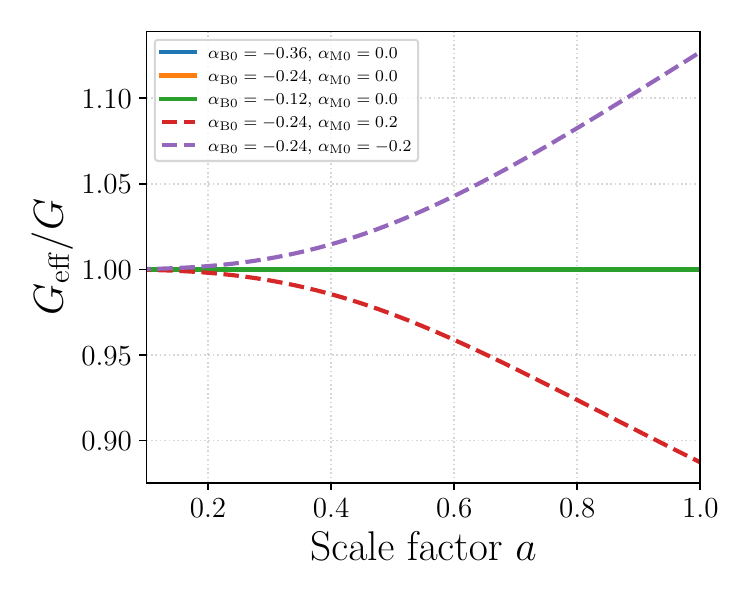}
      \caption{The change in the gravitational coupling, $G_{\rm eff}/G$, for different EFT parameter combinations, plotted against the scale factor $a$. The dashed lines show cases with varying $\amz$ compared to the green solid line. The blue and orange lines overlap with the green one because $\amz=0$ for all three lines.}
         \label{fig:geff}
   \end{figure}

With this time-dependence, we can further calculate the time-dependent Planck mass. From \citet{cusin}, we have \begin{equation}
    \am = \frac{M_{*}^2 \dot{f}}{M^2 H}= \frac{ \mathrm{d} \ln f}{\mathrm{d} \ln a},
\end{equation} where $M_{*} = 1/\sqrt{8 \pi G}$ is the standard Planck mass ($G$ is Newton's gravitational constant) and $M$ is the modified Planck mass, defined by $M^2 = M_{*}^2 f$, with $f$ being the function that encodes the change in the Planck mass over time. Consequently, we can define a time-varying gravitational coupling as  \begin{equation}
    \frac{G_{\rm eff} (a)}{G} = \exp\left[{-\int^a_0 \frac{\am(a')}{a'} \mathrm{d} a'}\right].
\end{equation} The exact functional form depends on the parameterisation chosen for the time-evolution of $\am$. Using the form given by Eqs.~(\ref{eq:alpha-a}) and for a $\Lambda$CDM background, we obtain \begin{equation}
    \frac{G_{\rm eff} (a)}{G} = [\Omega_m (a)]^{\amz / [3(1 - \Omega_{\rm m0})]},
\end{equation} where $\amz$ is the value of $\am$ at $z=0$. The evolution of $G_{\rm eff}/G$ versus scale factor is shown in Fig.~\ref{fig:geff}. The value stays at unity for $\amz=0$, increases to a value greater than 1 today for $\amz<0$, and goes to a value under 1 today for $\amz>0$. 

\subsection{Linearised equations}

Ignoring the terms with three or more copies of the fields in the Lagrangian (considering only $S_{\rm m} + S^{(2)}_{\rm g}$ from the action), the screening term vanishes and the resulting set of equations can be rearranged to obtain three decoupled linearised equations for the evolution of the three fields. This process is equivalent to ignoring the screening term in Eq.~(\ref{eq:chi_prelim}) and isolating the Laplacians of the three fields. This linearisation
yields three Poisson-like equations, one for each of $\varphi_a$: \begin{subequations} \label{eq:linearised}
\begin{align}
\label{eq:phi_lin}
\nabla^2 \Psi &= \frac{\bar{\rho}_{\rm m} a^2}{2 M^2}\, \mu_\Psi\, \delta, 
&\quad \mu_\Psi \equiv A^{-1}_{11} = 1 + \frac{\xi^2}{\nu}, \\[2pt] 
\nabla^2 \Phi &= \frac{\bar{\rho}_{\rm m} a^2}{2 M^2}\, \mu_\Phi\, \delta, 
&\quad \mu_\Phi \equiv A^{-1}_{12} = 1 + \frac{\xi \alpha_{\rm B}}{\nu}, \\[2pt]
\nabla^2 \chi &= \frac{\bar{\rho}_{\rm m} a^2}{2 M^2}\, \mu_\chi\, \delta, 
&\quad \mu_\chi \equiv A^{-1}_{13} = \frac{\xi}{\nu},
\end{align}
\end{subequations} where we have defined \begin{equation}
    \xi = \alpha_{\rm B} - \alpha_{\rm M},
\qquad
\nu = -\mathcal{C}_2 + \alpha_{\rm B}\,(2\am - \ab).
\end{equation}

Without modifications to gravity, we have $\mu_\Psi = 1$, $\mu_\Phi = 1$, and $\mu_\chi=0$, so that the equation for $\chi$ vanishes and those for $\Psi$ and $\Phi$ reduce to the standard Newtonian Poisson equation in an expanding universe, such that $\Psi = \Phi$ at all times. In the presence of the scalar field, the equations for $\Psi$ and $\Phi$ have different prefactors and the two fields evolve differently. Also note that $\nu$ is proportional to the product of the propagation speed of scalar fluctuations and the kinetic energy of the scalar mode, both of which should remain positive to avoid instabilities \citep{Cusin18_EFTofLSS}. Consequently, we require any viable model in this class to satisfy $\nu > 0$.

\begin{figure}[h!]
   \centering
   \includegraphics[width=\hsize]{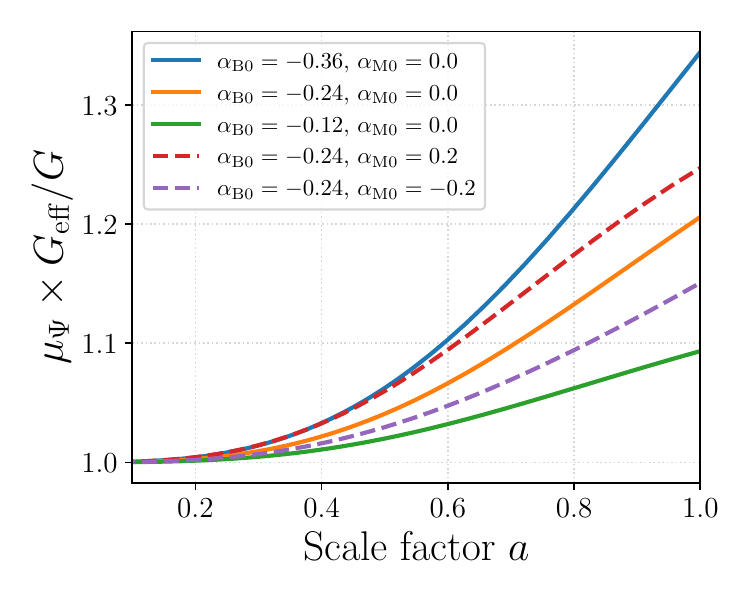}
      \caption{The prefactor $\mu_\Psi \times G_{\rm eff}/G$ for the Poisson-like linearised equation for the evolution of the gravitational potential $\Psi$ for different EFT parameter combinations plotted against the scale factor $a$. Continuous lines correspond to a variation of $\alpha_{\rm B0}$ from $-0.36$ (blue) to $-0.24$ (orange) and $-0.12$ (green). Dashed lines correspond to a variation of $\alpha_{\rm M0}$ between 0.2 (red) and -0.2 (purple).}
         \label{fig:muphi}
   \end{figure}

\subsection{Linear growth of the density field}

To facilitate comparisons of our simulation results with theory, here we derive the linear growth factor for perturbation modes deep inside the horizon in cosmologies with MG, in the EFTofDE formalism. Combining the continuity and Euler equations for small dark matter perturbations ($\delta \ll 1$) 
along with the modified linearised Poisson equation, we obtain 
\begin{equation}
\delta''(a)
+ \left[
\frac{3}{a}
+ \frac{\mathrm{d}\ln H}{\mathrm{d}a}
\right]\delta'(a)
- \frac{3}{2a^2}
\,\Omega_{\rm m}(a) \frac{G_{\rm eff} (a)}{G}\,
\mu_\Psi(a)\,
\delta(a)
= 0,
\label{eq:delta_mod_a}
\end{equation} where primes denote $\mathrm{d}/\mathrm{d}a$. This equation reduces to the standard linear growth equation for $\mu_\Psi = 1$ and $G_{\rm eff} = G$ ($M = M_{*}$). The growing mode solution for this equation yields the modified growth factor $D_{+}^{\rm EFT}(a)$. The factor $\mu_\Psi \times G_{\rm eff} / G$ quantifies the deviation of the linearised equation for $\Psi$ from standard GR. Figure \ref{fig:muphi} shows the time-evolution of this combined factor for various combinations of EFT parameters.

\begin{figure}[h!]
   \centering
   \includegraphics[width=\hsize]{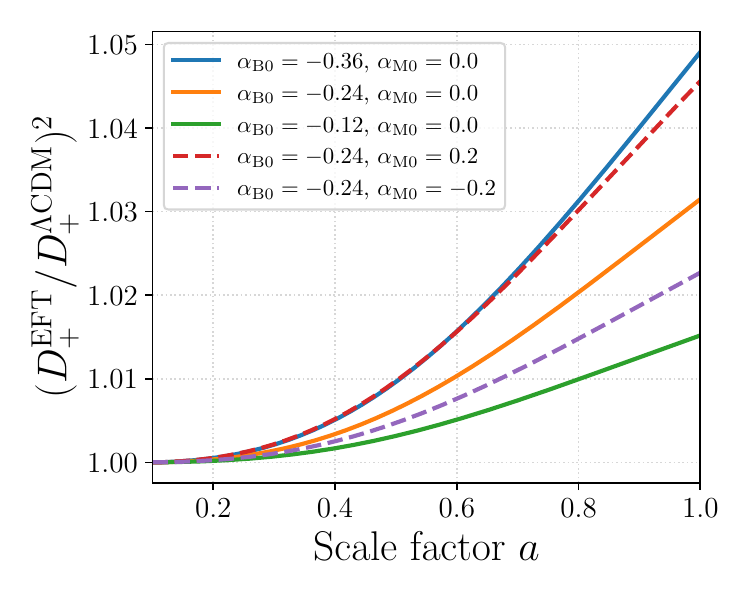}
      \caption{The squared ratios of the growth factors $D_+$ in EFT cosmologies to standard $\Lambda$CDM for different EFT parameter combinations, plotted against scale factor $a$. The color coding and the line style are the same as in Fig.~\ref{fig:muphi}.   }
         \label{fig:dplus}
   \end{figure}

Figure \ref{fig:dplus} shows the ratios of $(D_{+}^{\rm EFT}/ D_{+}^{\Lambda \rm CDM})^2$ for various $\abz$ and $\amz$ values as functions of scale factor $a$. All the curves have $\Omega_{\rm m0} = 0.3$. The solid curves show EFT scenarios with $\amz = 0$ and different $\abz$: increasing the magnitude of $\abz$ increases the large-scale boost. 

\subsection{Static spherically symmetric mass distribution}
\label{sec:sph-analytical}

The analytical solutions for the three fields can be obtained in the cubic screening case with a spherically symmetric mass distribution. Here, we present the solutions, taken from \citet{cusin}, which will be used in Sec.~\ref{spherical-sol} to test our solver. We first define the scaled radial derivatives of the three fields: \begin{equation} \label{eq:spherical-defs}
x(a,r) \equiv \frac{1}{a^{2}H^{2}} \frac{\chi'}{r}, 
\,
y(a,r) \equiv \frac{1}{a^{2}H^{2}} \frac{\Psi'}{r}, 
\,
z(a,r) \equiv \frac{1}{a^{2}H^{2}} \frac{\Phi'}{r},
\end{equation} and also the normalised integrated mass terms \begin{equation} \label{eq:Ar}
    A(a,r) \equiv \frac{1}{2 M^2 H^2 r^3}\int_0^r \tilde{r}^2 \delta \rho_{\rm m} (a,\tilde{r}) \textrm{d} \tilde{r}.
\end{equation} Considering the cubic action and writing equations for $x, y, z$, one obtains
\begin{equation} \label{eq:analytical}
    \begin{aligned}
    x &= \frac{\nu - \sqrt{\nu^{2} - 2 A \mathcal{C}_{4} \xi}}{\mathcal{C}_{4}}, \\
    y &= A + (\ab - \am) x, \\
    z &= A + \ab x.
\end{aligned}
\end{equation}
Note that $A(r)$ includes the mass enclosed within a radius $r$. For standard gravity, we have $y = A \,(=z)$ and the equation for $y$ in Eqs.~(\ref{eq:spherical-defs}) reduces to the Poisson equation for the gravitational potential $\Psi$. For $A \ll 1$, keeping the terms linear in $A$, the linearised equations~(\ref{eq:linearised}) are recovered.

\begin{figure}
    \centering
    \includegraphics[width=\linewidth]{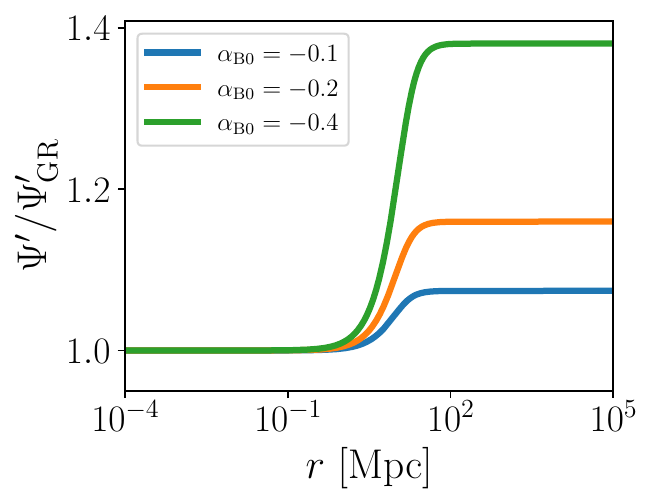}
    \caption{Ratio of the radial derivative of $\Psi$ (gravitational potential in the EFT case) to that of $\Psi_{\rm GR}$ (potential in standard GR) for a delta-function mass at $r=0$, plotted for different $\abz$ values (and $\alpha_{\rm M0}=0$ for all of the three cases).}
    \label{fig:spher-ana}
\end{figure}

Figure \ref{fig:spher-ana} shows the ratio $\Psi' / \Psi'_{\rm GR}$ plotted versus $r$, the distance from a delta-function mass of $10^{15} M_{\odot}$. The curves demonstrate EFT scenarios with three $\abz$ values and are evaluated with $\amz = 0$, $\Omega_{\rm m0} = 0.3$, and $H_0 = 70\,\rm km\,s^{-1}\,Mpc^{-1}$ at $a=1$. Increasing the magnitude of $\abz$ strengthens the coupling between the scalar field and the metric, increasing the ratio of gradients far away from the mass. As $r$ decreases, the effect of the Vainshtein screening mechanism reduces the ratios to unity, recovering the GR potential. 

\section{Methods}
\label{sec:methods}
In this section, we describe the implementation of the equations of motion for the scalar field, the calculation of the background EFT quantities, and the numerical methods used for $N$-body simulations in our modified versions of the \texttt{PySCo} and \texttt{ECOSMOG-CVG} codes.

The scalar field equation~(Eq.~\ref{eq:chi}) is elliptical and non-linear, which precludes the usage of Fourier-space and tree-based solution methods. We choose \texttt{PySCo} and \texttt{ECOSMOG-CVG} for their ability to solve differential equations on the grid using a combination of iterative solvers and multigrid schemes. Moreover, \texttt{PySCo}'s modular structure allows for easy modifications and its speed allows for a quick exploration of the EFTofDE parameter space. The user can conduct hundreds of fast PM-only $N$-body simulations for a preliminary assessment of the parameter space, and subsequently use the \texttt{ECOSMOG-CVG}-based code for simulating structure formation in highly-dense regions with adaptive mesh refinement.

\subsection{Code equations and units}

Both \texttt{PySCo} and \texttt{ECOSMOG-CVG} use supercomoving units \citep{martel1998}, which are defined by the following conversions from physical units \begin{equation}
    \tilde{x} = \frac{x}{x_*}, \quad \mathrm{d}\tilde{t} = \frac{\mathrm{d}t}{t_*}, \quad \tilde{v} = \frac{v\, t_*}{x_*}, \quad \tilde{\Psi} = \Psi\, \frac{t_*^2}{x_*^2}, \quad \tilde{\rho} = \frac{\rho}{\rho_*},
\end{equation} where a tilde denotes simulation coordinates. The conversion factors are \begin{equation}
    x_* = a L_{\mathrm{box}}, \quad 
t_* = \frac{a^2}{H_0}, \quad 
\rho_* = \frac{\Omega_{\rm m} \rho_{\rm c}}{a^3},
\end{equation} where $L_{\mathrm{box}}$ is the simulation box size in comoving coordinates, $H_0$ is the Hubble constant and $\rho_{\rm c}$ is the critical density at $z=0$. In supercomoving coordinates, the standard Poisson equation becomes \begin{equation}
    \tilde{\nabla}^2 \tilde{\Psi} = \frac{3}{2} a \Omega_{\rm m} (\tilde{\rho} - 1).
\end{equation} These transformations can be applied to Eqs.~(\ref{eqs:phipsi}) to obtain their counterparts in simulation coordinates. We obtain \begin{subequations}
\begin{empheq}[box=\fbox]{align}\label{eq:phipsi_mod}
    \tilde{\nabla}^2 \tilde{\Psi} &=  \frac{3}{2} \frac{G_{\rm eff}}{G} \, a \Omega_{\rm m} (\tilde{\rho} - 1) + (\ab - \am)\tilde{\nabla}^2 \tilde{\chi}, \\
    \tilde{\nabla}^2 \tilde{\Phi} &=  \frac{3}{2} \frac{G_{\rm eff}}{G} \, a \Omega_{\rm m} (\tilde{\rho} - 1) + \ab \tilde{\nabla}^2 \tilde{\chi}.
\end{empheq}
\end{subequations} 
The equation for the scalar field $\chi$, given by Eq.~(\ref{eq:chi}), 
\begin{equation} \label{eq:chi_scm}
\boxed{
\begin{aligned}
    & (2 \ab \am - \ab^2 - \mathcal{C}_2)\tilde{\nabla}^2 \, \tilde{\chi} \\
    & + (\am - \ab) \frac{3}{2} \frac{G_{\rm eff}}{G} \, a \Omega_{\rm m} (\tilde{\rho} - 1) \\
    & - \frac{\mathcal{C}_4}{4E^2 a^4}\left[ (\tilde{\nabla}^2 \tilde{\chi})^2 - \tilde{\nabla}_i \tilde{\nabla}_j \tilde{\chi} \, \tilde{\nabla}^i \tilde{\nabla}^j \tilde{\chi} \right] = 0
\end{aligned}
}
\end{equation} where $E(a) = H(a) / H_0$, the dimensionless Hubble parameter. Note that the for an $N$-body simulation, only $\Psi$ is required to move the particles at each time step. For the model discussed in this work, our codes do not compute $\Phi$ on the grid. 

In the above equations, $\ab$ and $\am$ and thus the derived quantities $\mathcal{C}_2$ and $\mathcal{C}_4$ are time-varying. For convenience and brevity, we will omit the tildes hereafter. All calculations in this section are in supercomoving coordinates.

\subsection{Discretisation: solving on the grid}
\label{sec:discmethods}

Given the non-linear nature of the elliptical partial differential equations, we rely on iterative solvers such as Jacobi or Gauss-Seidel using a multigrid strategy to accelerate the convergence. In this section, we describe two methods to update the scalar field (i.e.~a single iteration). 

\subsubsection{Method I: solving for $\chi$}
\label{sec:method1}
The objective is to solve for the scalar field $\chi$ on the grid in an $N$-body simulation and then to solve for the modified $\Psi$ to obtain the acceleration of simulation particles. With the equation for the scalar field $\chi$ separated, we can first solve for $\chi$ given the density distribution on the simulation grid. We will first discretise Eq.~(\ref{eq:chi_scm}) on a three-dimensional square mesh with cell size $h$. 

We will first focus on the first two terms of Eq.~(\ref{eq:chi_scm}), leaving the non-linear third term with cross-derivatives for later. The first term is simply the Laplacian which when discretised becomes ($i$, $j$ and $k$ are the indices for the cell position in $x$, $y$, and $z$ cartesian coordinates) \begin{equation}
\begin{aligned}
    (2\ab\am - \ab^2 - \mathcal{C}_2)\, \times \\
\frac{
  \chi_{i+1,j,k} + \chi_{i-1,j,k}
+ \chi_{i,j+1,k} + \chi_{i,j-1,k}
+ \chi_{i,j,k+1} + \chi_{i,j,k-1}
- 6\chi_{i,j,k}
}{h^2},
\end{aligned}
\end{equation} while the second term is \begin{equation}
    (\am - \ab)\,\frac{3}{2}\,\frac{G_{\mathrm{eff}}(a)}{G}\,a\,\Omega_{\rm m}\,(\rho_{i,j,k} - 1).
\end{equation} 

We now turn our attention to the non-linear term. The expression in brackets in the third term in Eq.~(\ref{eq:chi_scm}) can be simplified to \begin{equation}
    \left[
2(\chi_{xx}\chi_{yy} + \chi_{yy}\chi_{zz} + \chi_{zz}\chi_{xx})
- 2(\chi_{xy}^2 + \chi_{xz}^2 + \chi_{yz}^2)
\right],
\end{equation} where subscripts denote spatial derivatives. To expand this, we use the following discrete forms for the various double spatial derivatives above \begin{equation}
\begin{aligned}
\chi_{xx} &= \frac{\chi_{i+1,j,k} - 2\chi_{i,j,k} + \chi_{i-1,j,k}}{h^2}, \\[6pt]
\chi_{yy} &= \frac{\chi_{i,j+1,k} - 2\chi_{i,j,k} + \chi_{i,j-1,k}}{h^2}, \\[6pt]
\chi_{zz} &= \frac{\chi_{i,j,k+1} - 2\chi_{i,j,k} + \chi_{i,j,k-1}}{h^2}, \\[6pt]
\chi_{xy} &= 
\frac{
\chi_{i+1,j+1,k} - \chi_{i+1,j-1,k}
- \chi_{i-1,j+1,k} + \chi_{i-1,j-1,k}
}{4h^2}, \\[6pt]
\chi_{xz} &= 
\frac{
\chi_{i+1,j,k+1} - \chi_{i+1,j,k-1}
- \chi_{i-1,j,k+1} + \chi_{i-1,j,k-1}
}{4h^2}, \\[6pt]
\chi_{yz} &= 
\frac{
\chi_{i,j+1,k+1} - \chi_{i,j+1,k-1}
- \chi_{i,j-1,k+1} + \chi_{i,j-1,k-1}
}{4h^2}.
\end{aligned}
\end{equation} Using the above stencils, the final discretised equation for $\chi$ can be written as \begin{equation} \boxed{
    P \chi_{i,j,k}^2 + Q \chi_{i,j,k} + R = 0,}
\label{eq:chi_discrete}
\end{equation} where \begin{equation} 
\begin{aligned}
P &= -\,\frac{6\,\mathcal{C}_4}{h^4\,a^4 E^2}, \\[8pt]
Q &= -\,\frac{6(2\ab\am - \ab^2 - \mathcal{C}_2)}{h^2}
+ \frac{2\,\mathcal{C}_4}{h^4\,a^4 E^2}\,S_{i,j,k}, \\[8pt]
R &= (\am - \ab)\,b_{i,j,k}
+ \frac{(2\ab\am - \ab^2 - \mathcal{C}_2)}{h^2}\,S_{i,j,k}
- \frac{\mathcal{C}_4}{4\,a^4 E^2}\,q^{(2)}_{i,j,k},
\end{aligned}
\end{equation} with \begin{equation}
    S_{i,j,k} =
\chi_{i+1,j,k} + \chi_{i-1,j,k}
+ \chi_{i,j+1,k} + \chi_{i,j-1,k}
+ \chi_{i,j,k+1} + \chi_{i,j,k-1},
\end{equation} \begin{equation}
    b_{i,j,k} = \frac{3}{2} \frac{G_{\rm eff}}{G} (\rho_{i,j,k} - 1),
\end{equation} \begin{equation}
\begin{aligned}
q^{(2)}_{i,j,k}
&= 
\frac{2}{h^4}
\Big[
(\chi_{i+1,j,k} - 2\chi_{i,j,k} + \chi_{i-1,j,k})
(\chi_{i,j+1,k} - 2\chi_{i,j,k} + \chi_{i,j-1,k}) \\
&\quad+
(\chi_{i,j+1,k} - 2\chi_{i,j,k} + \chi_{i,j-1,k})
(\chi_{i,j,k+1} - 2\chi_{i,j,k} + \chi_{i,j,k-1}) \\
&\quad+
(\chi_{i,j,k+1} - 2\chi_{i,j,k} + \chi_{i,j,k-1})
(\chi_{i+1,j,k} - 2\chi_{i,j,k} + \chi_{i-1,j,k})
\Big] \\[6pt]
&\quad
-\frac{2}{(4h^2)^2}
\Big[
(\chi_{i+1,j+1,k} - \chi_{i+1,j-1,k} - \chi_{i-1,j+1,k} + \chi_{i-1,j-1,k})^2 \\
&\qquad+
(\chi_{i+1,j,k+1} - \chi_{i+1,j,k-1} - \chi_{i-1,j,k+1} + \chi_{i-1,j,k-1})^2 \\
&\qquad+
(\chi_{i,j+1,k+1} - \chi_{i,j+1,k-1} - \chi_{i,j-1,k+1} + \chi_{i,j-1,k-1})^2
\Big].
\end{aligned}
\end{equation}

The solution for $\chi$ at a grid point can be obtained using the quadratic formula \begin{equation} \label{eq:quadratic}
    \chi_{i,j,k} = \frac{-Q - \sqrt{Q^2 - 4PR}}{2P}.
\end{equation} We choose the quadratic solution with a minus sign to match the linearized limit of the spherically symmetric solution for small masses.

\subsubsection{Method II: operator splitting}

Eq.~(\ref{eq:chi_scm}) can also be discretised to obtain a quadratic equation in $[\nabla^2 \chi]_{i,j,k}$. Numerical solution techniques like Method I described above can suffer from poor convergence in high-density regions; the operator splitting method \citep{llinares} can address these convergence issues. We explain the technique here. We are interested in solving an equation of the form \begin{equation} \label{eq:opsplitform} 
    \nabla^2 \chi + A \left[ (\nabla^2 \chi)^2 - (\nabla_l \nabla_m \chi)(\nabla^l \nabla^m \chi) \right] = B \bar{\rho} \delta,
\end{equation} where $A$ and $B$ are time-dependent coefficients independent of the field values. Let us recast this equation as a quadratic equation for $\nabla^2 \chi$, written as \begin{equation}
    (\nabla^2 \chi)^2 + A^{-1} \nabla^2 \chi - \Sigma = 0,
\end{equation} where $\Sigma = (\nabla_l \nabla_m \chi)(\nabla^l \nabla^m \chi) + (B/A) \bar{\rho} \delta $. The explicit solution for this quadratic equation is \begin{equation}
    \frac{-A^{-1} \pm \sqrt{A^{-2} + 4 \Sigma}}{2}.
\end{equation} However, this solution contains $\nabla^2 \chi$ inside $\Sigma$, which itself contains the value of the central cell when discretised. This can lead to slower convergence. To ameliorate this, we can rewrite the above equation as \begin{equation}
    (1 - w)(\nabla^2 \chi)^2 + A^{-1} \nabla^2 \chi - \bar{\Sigma} = 0,
\end{equation} where $\bar{\Sigma} = \Sigma - w (\nabla^2 \chi)^2$. Now, decomposing the $\nabla_l \nabla_m \chi$ term into its traceless and trace parts, we have \begin{equation}
    \nabla_l \nabla_m \chi = \frac{1}{3} \gamma_{lm} \nabla^2 \chi + \hat{\nabla}_l \hat{\nabla}_m \chi, 
\end{equation} where $\gamma_{lm}$ is the identity operator. Squaring the above expression and using the orthogonality of the trace and traceless parts, we obtain \begin{equation}
    (\nabla_l \nabla_m \chi)^2 = \frac{1}{3} (\nabla^2 \chi)^2 + (\hat{\nabla}_l \hat{\nabla}_m \chi)^2. 
\end{equation} Substituting this into the expression for $\bar{\Sigma}$, we have \begin{equation}
    \bar{\Sigma} = \left( \frac{1}{3} - w \right) [\nabla^2 \chi]^2 + (\hat{\nabla}_l \hat{\nabla}_m \chi)^2 + \frac{B}{A} \bar{\rho} \delta.
\end{equation}Now, for $w = 1/3$, the $(\nabla^2 \chi)^2$ term cancels out from $\bar{\Sigma}$. As a result, after discretisation, the quadratic solution depends only on the neighbours of the current grid cell. This method has already been implemented in \textsc{ECOSMOG} \citep{Li:2013nua,Barreira:2013eea} and been found to improve convergence of the multigrid solvers for such equations \citep{Winther:2015wla, Becker:2020azq}.

Applying this technique, Eq.~(\ref{eq:opsplitform}) can be rewritten as \begin{equation} \label{eq:opsplit} \boxed{
    \nabla^2 \chi + A\left[ \frac{2}{3} (\nabla^2 \chi)^2 - (\hat{\nabla}_l \hat{\nabla}_m \chi)^2 \right] = B \bar{\rho} \delta,}
\end{equation} which can then be solved as a quadratic equation for $\nabla^2 \chi$ on the grid. 

To obtain the field value $\chi_{i,j,k}$ at each grid point, a Newton{--}Gauss{--}Seidel iterative solver is utilised. At the start of each iterative sweep (red or black), the numerical $\tilde{L} \equiv [\nabla^2 \chi]_{i,j,k}$ is computed at the grid point $\{i,j,k\}$ using the neighbours of the cell using the standard Laplacian stencil. Next, the quadratic solution Laplacian is computed by solving Eq.~(\ref{eq:opsplit}) and its mean subtracted to yield $\bar{L}$. Finally, a Newton update is performed on $\chi_{i,j,k}$: \begin{equation}
    \chi_{i,j,k}^{\rm new} = \chi_{i,j,k}^{\rm old} - \frac{\tilde{L} - \bar{L}}{(-6/h^2)}.
\end{equation}

\subsection{A note about the specific treatment of voids}

Both the methods described in Section~\ref{sec:discmethods} obtain the solution to a quadratic equation: for $\chi_{i,j,k}$ in Method I and for $[\nabla^2 \chi]_{i,j,k}$ in Method II. The discriminants for these quadratic equations can become negative in underdense regions. As a consequence, there are no real solutions to the scalar field in these cells. 

Assuming the mass $\bar{\rho}_{\rm m} \delta$ assigned to a grid cell is spherically symmetric, we can derive the solution for the scalar field. Looking at the expression for $x$ in Eqs.~(\ref{eq:analytical}), the condition for the existence of real solutions is \begin{equation}
    \nu^2 - 2A\mathcal{C}_4 \xi \geq 0,
\end{equation} where $A \propto \delta$ in the cell. For certain combinations of $\delta , \nu$, $\xi$ and $\mathcal{C}_4$, this condition is not satisfied, making the discriminant negative and resulting in a complex value of $\chi$. The condition fails for underdense cells ($\delta \lesssim -0.5$) at late times ($a \gtrsim 0.5$), with the exact thresholds depending on the values of $\abz$ and $\amz$. 

Recent work by \citet{Moretti:2026axy} used this condition for the nonexistence of a real solution to the scalar field in voids to constrain the parameter space of MG theories. Earlier work by \citet{Winther_QSA} also investigated this issue and concluded that the terms ignored by the quasi-static approximation cannot address the issue of the breakdown of the scalar field in voids at late times, arguing that this was a true instability of the model. However, some users may want to consider the cubic EFTofDE model as an effective and approximate version of a more generic unknown theory that behaves similarly to cubic screening in overdense regions and is well-defined in voids. 

As a consequence, to address the issue of these voids in our simulations, we follow the ad-hoc workarounds reported in earlier simulation-based works on similar models by \citet{Barreira:2013eea} and \citet{Becker:2020azq}. For underdense cells with no real scalar field solution, we set the discriminant of the field equation to a very low value close to zero. 
This choice has a negligible impact on the matter power spectrum because the non-linear scales in the simulation are dominated by overdensities. The user can choose to constrain the parameter space of $\{\abz,\amz\}$ using the methodology in \citet{Moretti:2026axy} before running simulations. Employing the saturation mechanism described above, the codes presented in this work are nevertheless able to run simulations for models where voids present pathologies with imaginary field solutions.

\subsection{Iterative and multigrid methods}

As explained at the beginning of this section, we employ grid-based methods to solve the scalar field equation owing to its elliptical and non-linear nature. We use the iterative solvers in both our codes, because a closed-form solution can be obtained for both $\chi_{i,j,k}$ and $[\nabla^2 \chi]_{i,j,k}$. The Jacobi iteration method updates the value of the field in the current cell using neighbouring values from the old iteration; the whole grid of cells is frozen to the old values for each update. In contrast, the Gauss--Seidel method updates the value of each cell ``in-place'' using the most recently updated values of the neighbours. We employed the Jacobi method in the field solver in \texttt{PySCo-EFT} and the Gauss--Seidel method in \texttt{ECOSMOG}.  

Because the Gauss-Seidel and Jacobi methods are local -- they update each cell based on the immediate neighbours -- propagating information across a large grid can take a long time. To expedite the solution, a multigrid scheme is often used in conjunction with the iterative solvers. The multigrid method solves the desired equation on a cascading network of coarser grids, computing large-scale solutions and feeding the errors from the coarse solutions back to the finer grids as corrections. For more details on how the multigrid method is used in \texttt{RAMSES}, we refer the reader to \citet{guillet2011simple}. 

Due to the non-linearity of our equation for $\chi$, we employ the Full Approximation Storage (FAS) multigrid scheme in both the modified \texttt{PySCo} and \texttt{ECOSMOG} codes. Consider solving the discretised differential equation $\mathcal{L} u = f $ on a grid of size $h$. In standard linear multigrid schemes, an approximate solution $\tilde{u}_h$ is obtained for $\mathcal{L}_h u_h = f_h $. The residual $r_h = f_h - \mathcal{L}_h \tilde{u}_h$ is then transferred to the coarser grids of size $H$, with the correction $v$ being obtained by solving the error equation $\mathcal{L}_H v_H = r_H$. The correction is then prolongated and transferred to the finer grid to update the fine-level solution. This method is not applicable for a non-linear differential equation because it relies on the linearity of $\mathcal{L}$. The FAS solves the full equation on the coarser grids instead of the error equation, thus needing to store the full approximation of the solution at each level.

\subsection{Codes}
\label{sec:codes}
\subsubsection{\texttt{PySCo}-EFT}

\texttt{PySCo} is a Python-based particle-mesh $N$-body simulation code first presented in \citet{pysco}. It uses explicit type declarations and the \texttt{numba} library for fast computation and parallelisation. The code provides both multigrid and Fourier-based solving methods for the Poisson equation, in addition to support for parametrised gravity, $f(R)$ gravity, and modified Newtonian dynamics (MOND). \texttt{PySCo} uses supercomoving coordinates like \texttt{RAMSES} and its modular structure makes it suitable for adding support for new physics like the EFTofDE formalism described in this work. 

We made several modifications to \texttt{PySCo}, starting with adding modules to calculate $\ab$, $\am$, $G_{\rm eff} / G$, and quantities like $\mathcal{C}_2$ and $\mathcal{C}_4$ that are used in the scalar field equation. To solve for the scalar field $\chi$, we added a \texttt{numba}-supported quadratic equation solver for Eq.~(\ref{eq:chi_discrete}) to the multigrid logic implemented in \texttt{PySCo}.

The scalar field array is initialised using the linearised equation $\nabla^2 \chi_{i,j,k} = (3/2) \, a \, (G_{\rm eff} / G) \, \Omega_{\rm m} \mu_\chi \, (\delta_{i,j,k} - 1)$, assuming all cells have $\chi=0$ at the start of the simulation. For subsequent timesteps, the value of the field at the previous timestep is used as the initial guess. 

The field is first smoothed with a number of Jacobi iterations $N_{\rm pre}$ with under-relaxation such that $\chi_{i,j,k}^{N+1} = 0.6\chi_{i,j,k}^{*N} + 0.4\chi_{i,j,k}^{N}$, where $\chi_{i,j,k}^*$ is the solution to the quadratic equation using the current values (step $N$) of the cell $i,j,k$ and its neighbours, and $\chi_{i,j,k}^{N+1}$ is the updated value. Subsequently, the V-cycle FAS multigrid algorithm is applied until a cycle reduces the residual error by a factor smaller than 2.5. Finally, the solution is smoothed with $N_{\rm post}$ cycles with the same under-relaxation.

As Eq.~(\ref{eq:phipsi_mod}) shows, the scalar field modifies the right-hand side of the Poisson equation. To incorporate this effect, either the source term can be modified, or the additional force resulting from $\chi$ can be added to the Newtonian force computed from the standard Poisson equation. To implement this fifth force, the modified force equation becomes \begin{equation}\label{eq:force}
    \textbf{F} = \frac{G_{\rm eff}}{G}\textbf{F}_{\rm Newton} + (\ab - \am) \nabla \chi.
\end{equation} For completeness, we tested a version modifying the Poisson equation by adding $(\ab - \am) \nabla^2 \chi$ to the right-hand side, which yielded similar results to the force addition method. Nevertheless, we adopt the latter method in keeping with \citet{pysco}. 

The final change to \texttt{PySCo-EFT} is the addition of the $G_{\rm eff} / G$ factor to the Newtonian Poisson equation to implement the time-variation of the gravitational coupling in the EFT formalism, as shown in Eq.~(\ref{eq:force}). The modified Poisson solver computes the gravitational potential with this factor included. 

We additionally implemented a linearised EFT option within the code, which solves the linearised Poisson-like Eq.~(\ref{eq:linearised}) $\Psi$, bypassing the need to compute the additional field.

\subsubsection{\texttt{ECOSMOG-EFT}}

In order to run $N$-body simulations of the EFTofDE with adaptive mesh refinement, we modified the \texttt{ECOSMOG-CVG} code which was first presented in \citet{Becker:2020azq}. \texttt{ECOSMOG-CVG} follows the conventions and code structure of \texttt{ECOSMOG} \citep{ecosmog} and solves for the transverse and longitudinal modes of the scalar field for the cubic vector Galileon (CVG) model. The equation for the longitudinal mode is similar in structure to our EFT scalar field Eq.~(\ref{eq:chi_scm}). As a result, modifying the \texttt{ECOSMOG-CVG} code is a natural choice for implementing our cubic screening EFT model within a \texttt{RAMSES}-like \citep{teyssier02} code. 

The original code implements the cubic vector Galileon, cubic scalar Galileon (CSG), quintessence-CDM (QCDM), and the sDGP models, all of which have background cosmological evolution differing from $\Lambda$CDM histories. We restored the background evolution in the code to standard $\Lambda$CDM, since the EFTofDE does not alter that. 

The code solves for the scalar field $\chi$ using the operator-splitting method, obtaining $\nabla^2 \chi$ by a combination of Gauss--Seidel smoothing with red-black ordering and the FAS multigrid V-cycle algorithm. An equation of the form of Eq.~(\ref{eq:opsplit}) is first solved to obtain the Laplacian, from which $\chi_{l,m,n}$ at each grid point is calculated using a Newton-Gauss-Seidel update. We modified the coefficients of the equation solved for the transverse mode of the Galileon field within the \texttt{ECOSMOG-CVG} code, which mirrors the form of Eq.~(\ref{eq:opsplit}). For the \texttt{ECOSMOG-EFT} version, we now have \begin{equation}
    A \;=\; -\,\frac{\mathcal{C}_4}{4E^2a^4\left(2\ab\am-\ab^2-\mathcal{C}_2\right)}\,,
    \end{equation} and \begin{equation}
        B \;=\; -\,\frac{\left(\am-\ab\right)\dfrac{3}{2}\dfrac{G_{\rm eff}}{G}\,a\,\Omega_{\rm m}}
{2\ab\am-\ab^2-\mathcal{C}_2}\,.
    \end{equation}

With the scalar field $\chi$ on the grid thus computed, the total force on the particles is calculated as the sum of the Newtonian and the fifth force resulting from $\chi$, following the force addition method used in \texttt{PySCo-EFT}. We use the standard five-point gradient stencil implemented in \texttt{ECOSMOG} \citep{ecosmog} to compute the gradient of $\chi$ for the force addition. The $ (\ab - \am)\nabla \chi$ contribution is added to the Newtonian gravitational force, which itself is multiplied by a $G_{\rm eff}/ G$ prefactor. 

\subsection{Numerical setups and $N$-body simulations}

\subsubsection{Static spherically symmetric configuration}
\label{sec:sphsym_methods}

As the first test of our discretised equations, we implemented the additional field solver on a fixed grid with a spherically symmetric mass distribution. We wrote a \texttt{numba}-accelerated Python-based code with a numerical solver that employs Jacobi iterations to obtain the scalar field using Method I (described in Sec.~\ref{sec:method1}) We assume a static Universe with $H_0 = 70\,\rm km\,s^{-1}\,Mpc^{-1}$, $\Omega_{\rm m0} = 0.3$, $\abz = -0.5$, $\amz$ = 0 and $\dot{\alpha}_{\rm B} = 0$. We used a cubic grid of side length $3.2$\,Mpc at $a=1$, in which we set up an isothermal mass distribution with $\rho \propto r^{-2}$ and total mass $10^{12}\,M_{\odot}$ truncated at $r = 1$\,Mpc. We ran the solver for boxes with two resolutions, with grid sizes of 0.05 ($64^3$ cells) and 0.10\,Mpc ($32^3$ cells).

The mass term $A(r)$ was computed for this distribution using Eq.~(\ref{eq:Ar}) and the analytical solutions for the radial derivatives of $\Psi$, $\Phi$, and $\chi$ were obtained using Eqs.~(\ref{eq:analytical}). These derivatives were then numerically integrated to compute the field values as a function of radius, considering all potentials to be zero at $r = \infty$. We used the analytical solutions thus obtained to set the boundary conditions on the faces of the cube. 

To solve for $\chi$, we set its value in each grid cell with Eq.~(\ref{eq:quadratic}) and updated the grid using the Jacobi iteration method. With $\chi$ computed, we then solved the discretised versions of Eqs.~(\ref{eq:phipsi_mod}) using Jacobi iterations, with the numerically differentiated Laplacian of the solution for $\chi$ added to the right-hand side. The results are presented in Sec.~\ref{spherical-sol}.

\subsubsection{Validation test simulation suite}
\label{sec:validation_methods}
With the changes implemented in \texttt{PySCo} and \texttt{ECOSMOG-CVG}, we first conducted a validation test to confirm the agreement between the simulation results from the two codes. Initial conditions were generated using second-order Lagrangian perturbation theory (2LPT) with a modified\footnote{The standard \texttt{MPGRAFIC} was modified to optimise it to run on a large number of processors. In addition, this version includes the ability to generate initial conditions (ICs) using second-order Lagragian perturbation theory. It was used for instance in \citet{Rasera22_RayGal}.} version of the \texttt{MPGRAFIC} code \citep{mpgrafic} for a box of side $L_{\rm box} = 328.125 \, h^{-1}\,\rm{Mpc}$ with 256$^3$ particles. The ICs were generated by scaling a linear power spectrum computed by the \texttt{CAMB} code \citep{camb} back to a redshift where the root-mean-squared matter fluctuations at the grid level were $\sigma_{\rm grid,ini} = 0.05$; this corresponded to a starting redshift of $z_{\rm ini} = 56.88$. The cosmological parameters used were from the Planck 2018 data \citep{Planck:2018vyg}, chosen to match the ``LCDM'' version of the RayGal Simulation Suite \citep{Breton:2018wzk,Rasera22_RayGal}; Table \ref{tab:validation_params} lists them along with the numerical parameters used for the runs. 

For each code, we ran an EFTofDE simulation and a $\Lambda$CDM simulation with the same cosmological background (except $\abz = \amz = 0$ in the $\Lambda$CDM run). We also ran the linearised version of the EFTofDE simulation. For our work, we used the same ICs for the EFT and $\Lambda$CDM simulation runs to isolate the effect of the MG, keeping the initial conditions fixed between the standard and non-standard gravity scenarios. We follow the same convention as earlier works by providing $\sigma_8^{\rm \Lambda CDM}$ i.e. $\sigma_8$ for $\abz=\amz=0$ and a $\Lambda$CDM linear growth factor with otherwise identical cosmological parameters as the ones of the modified gravity cosmology; this is equivalent to providing the normalisation $A_s$).

All the simulations, in this section and thereafter, were run using the Cloud-in-Cell density assignment scheme. CIC-based simulations with the \texttt{ECOSMOG} code have been shown to demonstrate good agreement with other MG simulation codes, in both the matter power spectrum and the halo mass function \citep{Winther:2015wla}.

The \texttt{ECOSMOG-EFT} run was conducted as a particle-mesh only simulation with no adaptive mesh refinement levels (to compare with the PM-only \texttt{PySCo-EFT}). The ICs were converted to an HDF5 file readable by \texttt{PySCo-EFT}, which was then used to run the \texttt{PySCo-EFT} simulations. For both codes, we used $N_{\rm pre} = N_{\rm post} = 6$ for the multigrid solver for the additional field. However, as mentioned above, \texttt{PySCo-EFT} solves for $\chi_{i,j,k}$, the value of the scalar field at each grid point, whereas \texttt{ECOSMOG-EFT} uses operator splitting to solve for $[\nabla^2 \chi]_{i,j,k}$ at each grid point. The results are shown in Sec.~\ref{sec:validation_results}. Additional simulations to test the numerical convergence of the results were also run on \texttt{ECOSMOG-EFT}; the setups for those are presented in Appendix~\ref{sec:convtests_methods}.

\begin{table}[h]
\centering
\caption{Validation test: parameters}
\label{tab:sim_params}
\begin{tabular}{lll}
\hline
\hline
\textbf{Parameter} & \textbf{Symbol} & \textbf{Value} \\ 
\hline
\multicolumn{3}{l}{\textit{Cosmological Parameters}} \\
\hline
Hubble parameter & $h$ & \phantom{$-$}0.720 \\
Matter density & $\Omega_{\rm m0}$ & \phantom{$-$}0.257 \\
DE density & $\Omega_{\Lambda}$ & \phantom{$-$}0.743 \\
Power spectrum normalization & $\sigma_8^{\rm \Lambda CDM}$ & \phantom{$-$}0.801 \\
Scalar spectral index & $n_{\rm s}$ & \phantom{$-$}0.963 \\
Braiding parameter & $\alpha_{\text{B0}}$ & $-0.240$ \\
Planck mass run rate & $\alpha_{\text{M0}}$ & \phantom{$-$}0.000 \\
\hline
\multicolumn{3}{l}{\textit{Numerical Parameters}} \\
\hline
Box size $[h^{-1}\,\mathrm{Mpc}]$ & $L_{\text{box}}$ & 328.125 \\
Initial redshift & $z_{\text{ini}}$ & \phantom{0}56.88 \\
Number of particles & $N_{\text{part}}$ & $256^3$ \\
Number of pre-steps & $N_{\text{pre}}$ & \phantom{0}\phantom{0}6 \\
Number of post-steps & $N_{\text{post}}$ & \phantom{0}\phantom{0}6 \\
\hline
\hline
\end{tabular} 
\label{tab:validation_params}
\end{table}

\subsubsection{Physical study simulation suite}
\label{sec:eft_methods}

Once the validation and convergence tests were run, we ran EFTofDE{--}$\Lambda$CDM simulation pairs for a set of EFTofDE parameters. We conduct two sets of tests: the first for $\amz = 0, \abz = \{-0.36, -0.24, -0.12\}$ and the second with $\abz = -0.24, \amz = \{0.20,0.00,-0.20\}$. These $\abz-\amz$ combinations are the same as shown in Fig.~\ref{fig:dplus}, chosen to illustrate the impact of the two parameters on the power spectrum boost $P_{\rm EFT} / P_{\Lambda \rm CDM}$. Following the method used for the validation and convergence tests, we used the \texttt{MPGRAFIC} code to generate the initial conditions via 2LPT and a linear power spectrum calculated by \texttt{CAMB}. The cosmological parameters were taken from the Planck 2018 best fit parameters\footnote{See table 6.7 in \url{https://wiki.cosmos.esa.int/planck-legacy-archive/images/4/43/Baseline_params_table_2018_68pc_v2.pdf}} with $m_\nu=0$; they are listed along with the numerical parameters in Table~\ref{tab:eftsims_params}. We used $N_{\rm pre} = N_{\rm post} = 6$ and 6 AMR levels for all these runs, with a comoving box size of $328.125\,h^{-1}\,\textrm{Mpc}$ and 512$^3$ particles. 

All power spectra presented in Appendix \ref{sec:convtests_results} and in this section were computed using the \texttt{powergrid} code \citep{mpgrafic} using Cloud-in-Cell density assignment. Power spectrum estimation based on Fast Fourier Transform (FFT) methods is known to lead to aliasing, which adds spurious power near the Nyquist wavenumber of the grid. This issue was addressed by computing the density on a grid that was much finer than the particle grid in the simulations and restricting the power spectrum plots to $k < k_{\rm Nyquist}/2$ for the power spectrum grids. The finer grid increases the Nyquist wavenumber to higher $k$, reducing the effect of aliasing for a larger range of $k$. The $k < k_{\rm Nyquist}/2$ cut eliminates the small-scale modes that are most affected by aliasing. These combined strategies have been shown to be effective at reducing aliasing much below the percent level \citep{alimi10}. Moreover, the results in this work are concerned with the power spectrum boost, which suppresses systematics on account of being a ratio \citep{ICS_emantis}. In future work, we plan to use more accurate power spectrum estimation techniques \citep{colombi09} like higher-order density assignment schemes and interlacing \citep{Sefusatti:2015aex}.

\begin{table}[h]
\centering
\caption{Physical study: parameters}
\begin{tabular}{lll}
\hline
\hline
\textbf{Parameter} & \textbf{Symbol} & \textbf{Value} \\ 
\hline
\multicolumn{3}{l}{\textit{Cosmological Parameters}} \\
\hline
Hubble parameter & $h$ & 0.6803 \\
Matter density & $\Omega_{\rm m0}$ & 0.3071 \\
DE density & $\Omega_{\Lambda}$ & 0.6929 \\
Power spectrum normalization & $\sigma_8^{\rm \Lambda CDM}$ & 0.8220 \\
Scalar spectral index & $n_{\rm s}$ & 0.9660 \\
Braiding parameters & $\abz$ & \{-0.36,-0.24,-0.12\} \\
Planck mass run rates & $\amz$ & \{0.20,0.00,-0.20\} \\
\hline
\multicolumn{3}{l}{\textit{Numerical Parameters}} \\
\hline
Box size $[h^{-1}\,\mathrm{Mpc}]$ & $L_{\text{box}}$ & 328.125 \\
Number of particles & $n_{\text{part}}$ & $512^3$ \\
Initial redshift & $z_{\text{ini}}$ & 47.71 \\
Pre-steps & $N_{\text{pre}}$ & 6 \\
Post-steps & $N_{\text{post}}$ & 6 \\
Refinement threshold & $m_{\rm ref}$ & 14 \\
\hline
\hline
\end{tabular} 
\label{tab:eftsims_params}
\end{table}

\section{Results}
\label{sec:results}

\subsection{Test of the iterative solver for a static spherically symmetric distribution} \label{spherical-sol}

\begin{figure*}[t]
    \centering
    \begin{subfigure}[t]{0.48\textwidth}
        \centering
        \includegraphics[width=\linewidth]{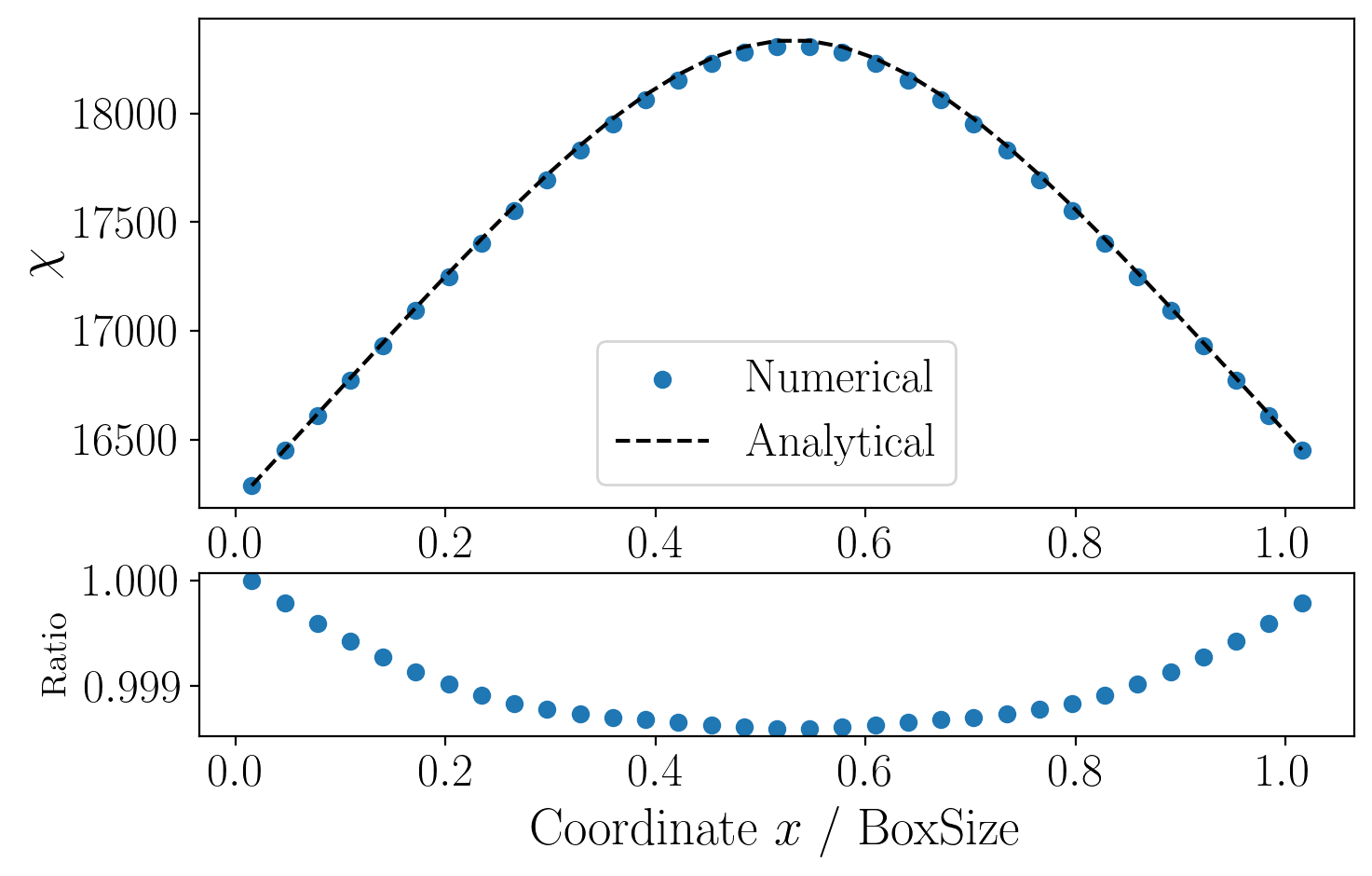}
        \caption{32$^3$ grid cells, $h = 0.1$ Mpc.}
        \label{fig:panel1}
    \end{subfigure}
    \hfill
    \begin{subfigure}[t]{0.48\textwidth}
        \centering
        \includegraphics[width=\linewidth]{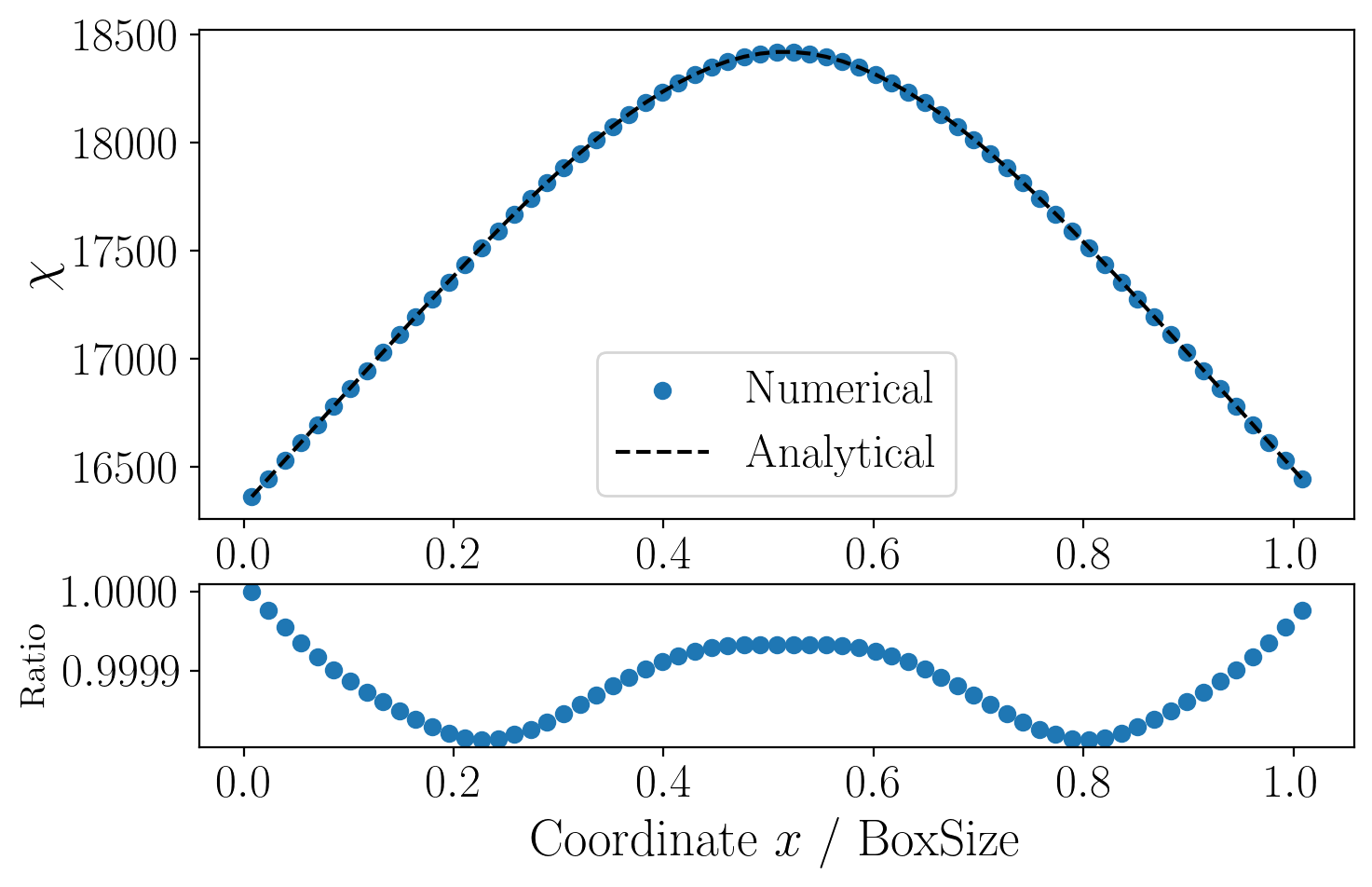}
        \caption{64$^3$ grid cells, $h=0.05$ Mpc.}
        \label{fig:panel2}
    \end{subfigure}
    \caption{Solutions for $\chi$ on a fixed grid at $a = 1$, with $\ab=-0.5$ and $\am = 0$. \textit{Top}: binned mean values of $\chi$ (dots) with the box viewed from the side; analytical solutions (dashed). \textit{Bottom}: ratio of the numerical to the analytical solution in each bin.}
    \label{fig:grids}
\end{figure*}

 Here, we present the results of the iterative solver for the additional scalar field $\chi$ with a static, spherically symmetric mass distribution (described in Sec.~\ref{sec:sphsym_methods}). Figure~\ref{fig:grids} shows the solutions for $\chi$ (dots) in the boxes viewed from a side-on angle, compared to the analytical solutions (dashed lines). The dots show the mean $\chi$ in bins of the $x$-coordinate. The left panel shows the results for a box with grid size $h = 0.1$\,Mpc and 32$^3$ cells, while the right panel is for a box with $h=0.05$\,Mpc and 64$^3$ cells. In the bottom panels of both plots, we show the ratio of the numerical to analytical solutions in each bin. The numerical solutions for $\chi$ match the analytical versions to within 0.2\,\% for both boxes. 
 
\begin{figure}[h!]
   \centering
   \includegraphics[width=\hsize]{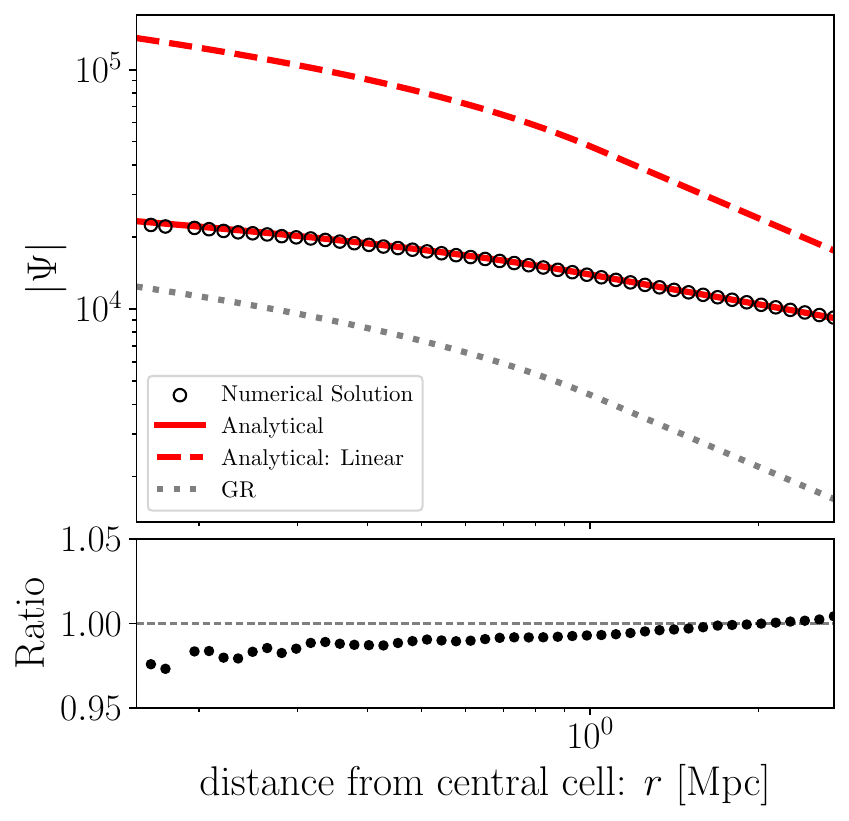}
      \caption{\textit{Top}: The absolute value of the gravitational potential $\Psi$ for the $64^3$ cell box (circles), binned by radius. The solid curve shows the analytical solution, the dashed curve shows the analytical solution ignoring the screening mechanism, and the dotted line shows $\Psi$ in a standard GR scenario. \textit{Bottom}: the ratio of the numerical solution to the analytical result (solid curve in the top panel).}
         \label{fig:allphi64}
   \end{figure}

 Furthermore, we display the numerical solutions of $\Psi$ (circles) versus the analytically obtained $\Psi$ (solid red) for the 64$^3$ cell box as a function of radius in Fig.~\ref{fig:allphi64}. The two agree to within 2.5\,\%, as shown in the lower panel of the same plot. The agreement remains within the 2\,\%  level down to 0.2~Mpc corresponding to 4 cells. Below this scale, we expect discretisation effects to play an important role. We would also like to note the lack of multigrid acceleration in the solver used for this specific test.
 
 The upper panel also plots the analytical $\Psi$ for the linearised EFT scenario, ignoring screening (dashed red). The solid curve is much below the dashed, showing that we are in the non-linear regime and Vainshtein screening reduces the magnitude of the potential. For comparison, we also plot the standard GR $\Psi$, which is much lower than the numerical solution and the solid red line, exhibiting how the scalar field-metric coupling enhances $\Psi$. Overall, this result demonstrates that the solver performs well at recovering the static spherically symmetric solutions, which are particularly relevant for the physics at the halo scale. Note that in this case, because $\am = 0$, we have $\Psi = \Phi$.  

\subsection{Validation test: \texttt{PySCo-EFT} vs \texttt{ECOSMOG-EFT} and convergence towards linear theory at large scales}
\label{sec:validation_results}

In this section, we show the results of the validation test comparing simulations from \texttt{PySCo-EFT} and \texttt{ECOSMOG-EFT} (particle-mesh), the methods for which were described in Sec.~\ref{sec:validation_methods}. All the power spectra used for the results in this subsection were computed using the \texttt{PK Library} within the \texttt{Pylians} python package \citep{Pylians} with a Cloud-in-Cell density assignment scheme. The \texttt{ECOSMOG-EFT} power spectra were scaled back to $z=0$ using the linear growth factors $D_+^{\rm EFT}$ and $D_+^{\Lambda \rm CDM}$, because \texttt{ECOSMOG-EFT} does not output snapshots at exactly specified times. 

\begin{figure}[h!]
   \centering
   \includegraphics[width=\hsize]{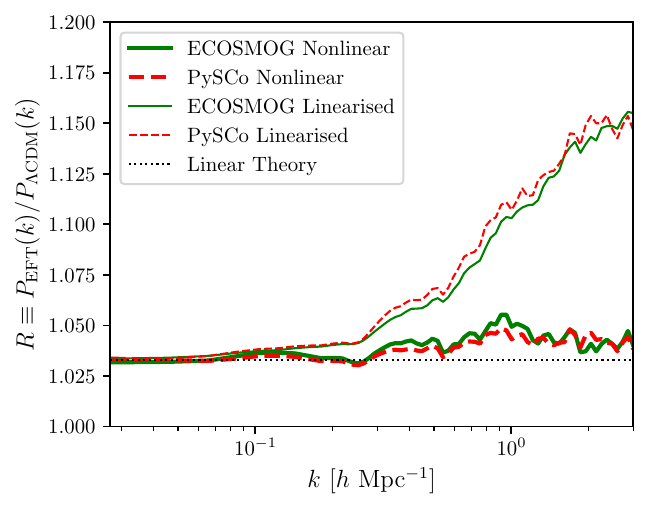}
      \caption{EFT boost $R(k) \equiv P_{\rm EFT} / P_{\Lambda \rm{CDM}}$ at $z = 0$ from the linearised (thin lines) and full non-linear (thick lines) simulations in \texttt{PySCo-EFT} (dashed) and \texttt{ECOSMOG-EFT} (solid). The black dotted line shows $(D_+^{\rm EFT} / D_+^{\Lambda \rm{CDM}})^2$ obtained from linear theory for this EFT scenario ($\abz = -0.24, \amz = 0$).}
         \label{fig:pysco-psrats}
   \end{figure}

Figure~\ref{fig:pysco-psrats} shows the boost \begin{equation}
    R(k) \equiv P_{\rm EFT} / P_{\Lambda \rm CDM},
\end{equation} the ratio between the EFT and $\Lambda$CDM power spectra for pairs of simulations. The solid green curves show the $R(k)$ from the simulations in \texttt{ECOSMOG-EFT}, while the dashed red curves show the boost from \texttt{PySCo-EFT} simulations. The thin lines show the boost from the linearised simulations, while the thick lines show those from the full non-linear simulations with Vainshtein screening. All the $R(k)$ curves have been smoothed with a one-dimensional Gaussian filter with a wavelength of $0.019\,h\,\rm Mpc^{-1}$.

The boosts show how both EFT power spectra are slightly higher than the $\Lambda$CDM version at all scales, with the non-linear and linear EFT curves converging at large scales (small $k$) and diverging from each other at small scales ($k \gtrsim 0.3 \,h\,\rm Mpc^{-1}$). In the linearised case, the Poisson equation is modified by a prefactor $\mu_\Psi$ which is greater than 1 in this scenario. This modification ``amplifies'' gravity, an effect which increases on smaller scales due to non-linear structure formation. Consequently, $R(k)$ keeps rising with increasing $k$. For the full non-linear $R(k)$, the Vainshtein screening mechanism {--} implemented via the cross-derivative terms in Eq.~(\ref{eq:chi_scm}) {--} inhibits the amplification of gravity on small scales, leading to $R(k)$ falling lower than the linearised version on small scales. 

Additionally, the black dotted line marks the ratio $(D_+^{\rm EFT} / D_+^{\Lambda \rm{CDM}} )^2$ obtained using the linear growth equation~(Eq.~\ref{eq:delta_mod_a}). Both the linear and non-linear $R(k)$ from both our codes are within 0.4\,\% of this linear $(D_+^{\rm EFT} / D_+^{\Lambda \rm{CDM}} )^2$ for $k < 0.05 \, h \, \rm{Mpc}^{-1}$, showing the agreement of the simulation results from both codes with linear theory at small wavenumbers. 

\begin{figure}[h!]
   \centering
   \includegraphics[width=\hsize]{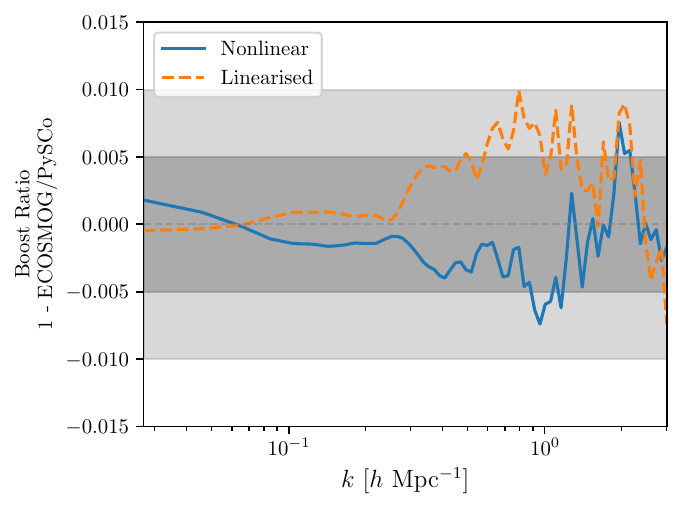}
      \caption{The relative error between the EFT boosts $R(k)$ obtained from \texttt{PySCo-EFT} and \texttt{ECOSMOG-EFT} (particle-mesh) at $z=0$. The solid line shows the error between the boosts for the non-linear simulations, while the dashed line shows the same for the linearised simulations. The grey bands show the 0.5\,\% and 1\,\% limits. Note that the solvers are different but the difference remains below the percent level for all scales $k\lesssim 3 \, h \, \rm{Mpc}^{-1}$.}
         \label{fig:pysco-pspercs}
   \end{figure}

In Fig.~\ref{fig:pysco-pspercs}, we plot the relative errors between the $R(k)$ values obtained from \texttt{PySCo-EFT} and \texttt{ECOSMOG-EFT}, with the solid curve showing the relative error between $R(k)$ from the non-linear simulations and the dashed curve showing the same from the linearised simulations. Both the curves are within 1\,\% for all the scales probed by the simulation boxes, displaying excellent agreement between the two codes despite two different methods being used for computing the scalar field. To complete these tests, we have performed a convergence study of the numerical parameters, with the results presented in Appendix~\ref{sec:convtests_results}.

\subsection{Physical study: impact of $\abz$ and $\amz$ on the matter power spectrum}

To highlight the impact of varying the EFTofDE parameters $\abz$ and $\amz$ on the matter power spectrum, we ran a suite of test simulations (described in Sec.~\ref{sec:eft_methods}) where the numerical parameters are chosen in the percent level converged region (see Appendix~\ref{sec:convtests_results}). In this section we describe those results. We make a few quantitative observations from these results, leaving a more detailed study of the EFTofDE using these tools to future work.

\begin{figure}
   \centering
   \includegraphics[width=\hsize]{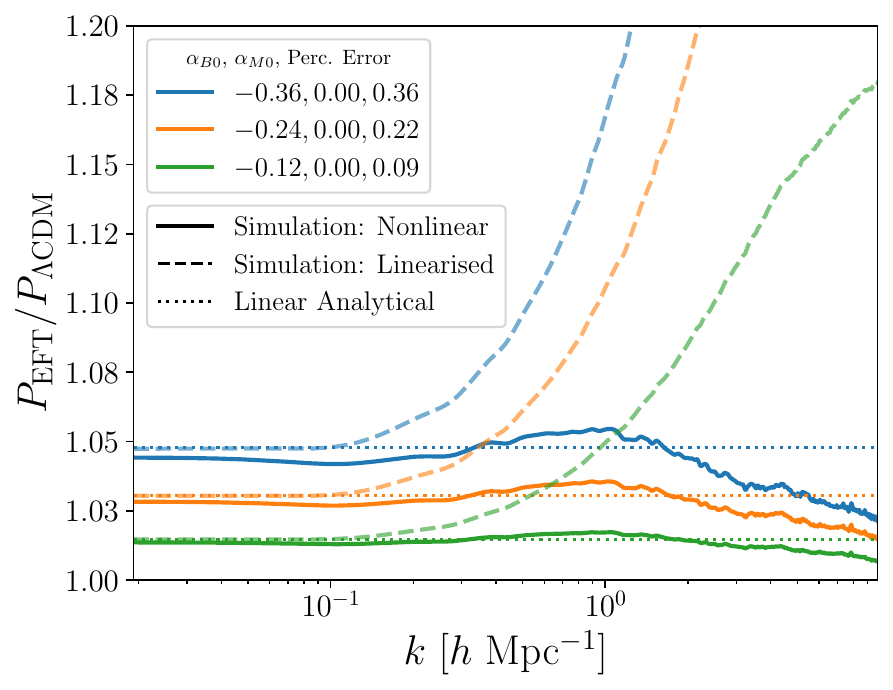}
   \caption{EFT boost $R(k) \equiv P_{\rm EFT} / P_{\Lambda \rm{CDM}}$ at $z=0$ for different values of $\alpha_{\rm B0}$. Solid curves show boosts from the full non-linear simulations, dashed curves show boosts from the linearised simulations, and dotted lines show the analytical predictions from linear theory. The percentage error shows the difference between the linear analytical prediction and the first $k$-bin of the non-linear simulation boost.}
   \label{fig:eft-ab0}
\end{figure}

Figure~\ref{fig:eft-ab0} shows the power spectrum boost $P_{\rm EFT} / P_{\Lambda \rm CDM}$ for $\abz = \{-0.36, -0.24, -0.12 \}$ and $\amz = 0$. The solid lines show the boost from the full non-linear simulations with the screening, while the dashed lines show the boost from their corresponding linearised versions. All the boosts have been smoothed with a 1D Gaussian filter with a $\sigma_{\rm filter} = 0.028 \, h \, \rm{Mpc}^{-1} \,  $ to generate smoother curves. We also plot the linear theory predictions $(D_+^{\rm EFT} / D_+^{\Lambda \rm CDM})^2$ for each case with dotted horizontal lines. 

Increasing the magnitude of $\abz$ increases the coupling strength between the scalar field and the metric. This is shown by the increasing large-scale (small $k$) values of the boost. In the quasi-linear regime ($0.03< k < 2 \, h \, \rm{Mpc}^{-1}$), the boost increases due to the non-linear growth rate.
In the non-linear regime, a comparison between the dashed and solid curves illustrates the impact of the Vainshtein screening: the linearised boost increases to very high values at small scales, but the screening reduces the boost to levels close to unity, restoring standard GR in highly dense regions. 

Both the non-linear and linearised curves are extremely close to the corresponding linear predictions (dotted lines) on large scales (linear regime). The legend of Fig.~\ref{fig:eft-ab0} shows the percentage error between the boost in the bin with smallest $k$ from the simulation and the linear prediction. The error remains lower than 0.4\,\% for $\abz = -0.36$. 

\begin{figure}[]
   \centering
   \includegraphics[width=\hsize]{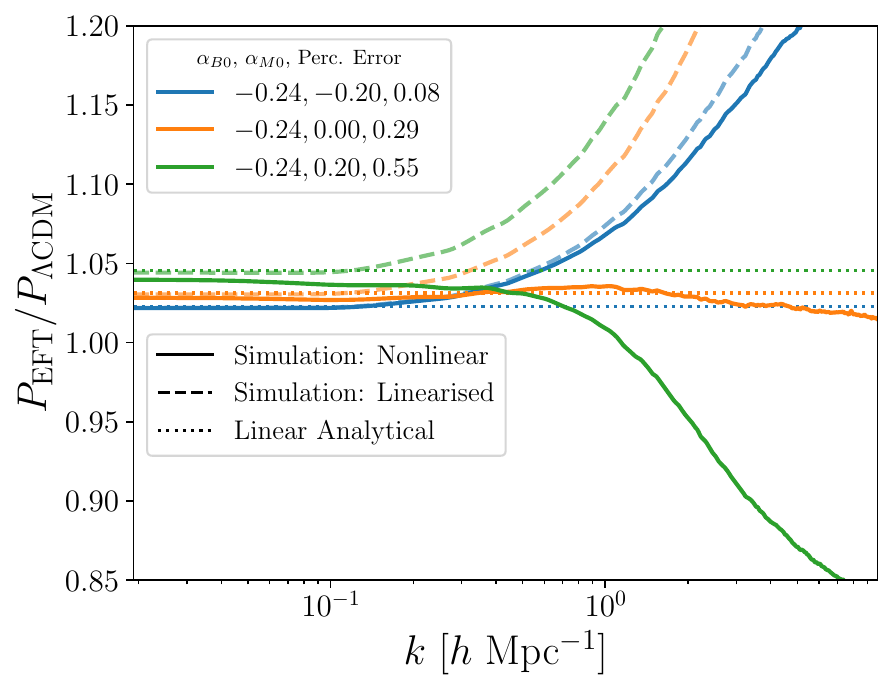}
   \caption{EFT boost $R(k) \equiv P_{\rm EFT} / P_{\Lambda \rm{CDM}}$ at $z=0$ for different values of $\alpha_{\rm M}$. Solid, dashed, and dotted curves represent the same categories as in Fig.~\ref{fig:eft-ab0}.}
   \label{fig:eft-am0}
\end{figure}

In Fig.~\ref{fig:eft-am0}, we show the boost curves, their linearised versions, and the analytical predictions for $\abz = -0.24$ and $\amz = \{-0.20,0,0.20\}$. The simulation results show excellent agreement with the analytical predictions on large scales, with the highest percentage error being under 0.6\%. We note that preliminary tests indicate this offset on linear scales is proportional to $\abz - \amz$; we conjecture that it results from the backreaction of the non-linear screening terms on the modified Poisson equation. We leave a deeper investigation of this offset for future work. 

Figure \ref{fig:eft-am0} also shows the three curves diverge at small scales; this is the result of $\am$ changing the gravitational coupling. A negative $\amz$ increases the coupling in time, strengthening the effect of gravity and enhancing small-scale structure formation. This results in the increasing small-scale boost for increasing $k$ in the case where $\amz=-0.2$. The opposite effect is seen for $\amz=0.2$. 

It is worth noting that the difference between the linearised and non-linear spectra increases as a function of the magnitude of $\mathcal{C}_4 \equiv -4\ab + 2\am$. This trend is consistent with the expectation from Eq.~\ref{eq:chi}. We also checked that if $\mathcal{C}_4\approx 0$ we obtain sub-percent level agreement between spectra from the linearised and non-linear runs. The often-used linearised approximation is therefore valid for cases in which the magnitude of $\mathcal{C}_4$ is small. However, the use of the linearised approximation where the screening is neglected should be avoided for large values of $\mathcal{C}_4$ since it can induce large errors (e.g. 1\% for $\mathcal{C}_4 |_{z = 0}=0.56$ at $k = 2\,h\,\rm Mpc^{-1}$, 15\% for $\mathcal{C}_4 |_{z=0}=0.96$ and 35\% for $\mathcal{C}_4|_{z=0}=1.36$ for the models considered).

\section{Conclusions}
\label{sec:conclusions}

In this work, we have presented two fast and accurate $N$-body codes, \texttt{PySCo-EFT} and \texttt{ECOSMOG-EFT}, for simulating cosmic structure formation in MG theories described by the Effective Field Theory of Dark Energy (EFTofDE) formalism. This formalism provides a compact method to describe a wide range of MG scenarios. While originally designed to address the linear regime, its extension to include non-linear terms in the scalar field equation \citep[see e.g.][]{cusin} makes the formalism able to incorporate the effects of screening, which are usually relevant in the non-linear regime of structure formation. While our approach is generic, in this work we consider a subset of Horndeski theories (in which one additional scalar field acts like the DE component driving the accelerated expansion of the Universe). Specifically we focus on a cubic Lagrangian featuring the Vainshtein screening mechanism. Accounting for the luminal speed of gravitational waves, the EFTofDE Lagrangian is parameterised by two numbers: $\abz$, the braiding parameter, which controls the coupling of the scalar field to the metric; and $\amz$, which defines the time-evolution of the Planck mass. For this work, we consider the time dependence of these parameters to be proportional to $1-\Omega_{\rm m}(a)$.

We implemented this parameterised EFTofDE model in two codes: \texttt{PySCo-EFT}, a fast Python-based particle mesh code based on \texttt{PySCo} \citep{pysco}, and \texttt{ECOSMOG-EFT}, a \texttt{RAMSES}-based code with adaptive mesh refinement based on \texttt{ECOSMOG-CVG} \citep{Becker:2020azq}. \texttt{PySCo-EFT} can be used for quick exploration of the parameter space, while \texttt{ECOSMOG-EFT} can be used for accurate simulations including the effects of MG in high density regions. In anticipation of plentiful and precise data from upcoming Stage-IV large-scale-structure surveys, these tools provide the ability to generate accurate predictions for constraining the $\alpha$-parameters that quantify the physical effects of MG theories. 

The codes presented in this paper use the Jacobi and Gauss--Seidel iterative methods coupled with multigrid schemes to calculate the additional scalar field $\chi$ on the grid, solving a non-linear elliptical partial differential equation (instead of a linear one for the Poisson equation). The fifth-force contribution proportional to $\nabla \chi$ is then added to the gravitational field of the particles. Given the non-linearity of the scalar field equation, the full  EFTofDE simulations are computationally intensive and cost about $10$ times the CPU time of the $\Lambda$CDM-only runs. We have also included a ``linearised'' EFT option in both codes that discounts the screening; these simulations enable a resource-light test of the theory on linear scales; they have the same computational cost as the $\Lambda$CDM versions.

We performed several validation tests spanning a range of scales. At small scales, we checked that an iterative Jacobi solver recovers static spherically symmetric solutions. At intermediate scales, we compared the results from the two $N$-body codes in a cosmological setup. The power spectrum boost $P_{\rm EFT} / P_{\Lambda \rm CDM}$ for both the linearised and non-linear simulation runs demonstrated sub-percent level agreement between \texttt{PySCo-EFT} and \texttt{ECOSMOG-EFT} for wavenumbers $k \leq 3\,h\,\rm Mpc^{-1}$. This agreement is notable, since the solvers are implemented differently in the two codes, computing the scalar field in \texttt{PySCo-EFT} and its Laplacian in \texttt{ECOSMOG-EFT}.  

Additionally, we conducted convergence tests for \texttt{ECOSMOG-EFT}, displaying two-percent-level robustness of the results to variations in multiple numerical parameters up to $k = 10\,h\,\rm Mpc^{-1}$ . These parameters included the mass resolution of the simulation, the box size, the pre- and post-smoothing cycles for the multigrid solver, the mass refinement threshold for triggering adaptive mesh levels, the initial redshift, and the residual threshold for the stopping criterion of the scalar field solver.

Finally, we also performed a physical study to highlight the impact of $\abz$ and $\amz$ on the matter power spectrum boost by conducting simulations for a few combinations of these parameters. Increasing the magnitude of the $\abz$ parameter enhanced the amplitude of the boost in the linear regime. However, the boost in the $\amz=0$ cases all approached unity at very small scales, indicating a recovery of standard general relativity in highly dense environments due to Vainshtein screening. On the contrary, the boost increased with increasing wavenumber in a non-physical way in simulations with the often-used linearised scalar field equations. Varying $\amz$ demonstrated significant impact on the boost in the non-linear regime: a smaller value of $\amz$ led to a higher value of the boost. At large (linear) scales, the power spectrum boosts obtained from the \texttt{ECOSMOG-EFT} simulations agree with the linear theory predictions to within 0.6 percent for the parameter combinations we tested, with a maximum error of 0.55\% for the largest magnitude of $\abz-\amz$ ($\abz=-0.24, \amz=0.20$). Overall, our tests show that the impact of screening on small scales becomes significant as the magnitude of $\mathcal{C}_4 \equiv -4\ab + 2\am$ increases; the linearised simulations are inaccurate in these cases and should be avoided.   

Several improvements to the \texttt{ECOSMOG-EFT} code are in preparation including a more generic time evolution of the~$\alpha$s. Additionally, the background is fixed to $\Lambda$CDM currently, which we plan to augment with the capability to work with a user-specified cosmological background via an input table of the cosmological history. These additions will allow the exploration of a much broader ranger of cosmological models in the future.

The tools presented in this work enable $N$-body simulations of MG models within the EFTofDE framework, focusing on the physically relevant parameters of such theories and enhancing our ability to generate precise predictions for observational constraints on a diverse range of scales. To the best of our knowledge, \texttt{PySCo-EFT} is the first Python-based code for $N$-body simulations for the EFTofDE, and \texttt{ECOSMOG-EFT} is the first AMR-based implementation of the EFTofDE with screening in the $\alpha$-basis. As part of upcoming work, we plan to use these tools for a detailed exploration of the EFTofDE parameter space, which will include generating 3D large-scale structure and lensing maps, power spectra, halo mass functions, and an emulator for the EFTofDE boost up to non-linear scales.


\begin{acknowledgements}
      We would like to thank Petter Taule for kindly providing EFT-based power spectrum predictions from his modified \texttt{hiclass} code. We also thank Filippo Vernizzi and Guilherme Brando for enlightening discussions about the theoretical aspects of the EFTofDE. We thank R.~Teyssier for making \texttt{RAMSES} public as well as C.~Becker, C.~Arnold, B.~Li and L.~Heisenberg for making \texttt{ECOSMOG-CVG} public. We would also like to thank Fabrice Roy for help in developing and optimising the codes, and Stephane Mene for help in maintaining the various computing clusters on which tests were conducted. 
      
      The authors acknowledge the support of the French Agence Nationale de la Recherche (ANR), under grant ANR-23-CE31-0010 (project ProGraceRay). This project was provided with computing HPC and storage resources by GENCI at TGCC thanks to the grant 2025-A0190402287 on the supercomputer Irene Joliot Curie's ROME partition. Part of the simulations in this work were conducted on the Infinity cluster at the Institut d’Astrophysique de Paris (IAP).
      EB is supported by the European Union’s Horizon Europe research and innovation programme under the Marie Sklodowska-Curie Postdoctoral Fellowship Programme, SMASH co-funded under the grant agreement No.~101081355. Disclaimer: Co-funded by the European Union. Views and opinions expressed are however those of the author(s) only and do not necessarily reflect those of the European Union or European Research Exacutive Agency. Neither the European Union nor the granting authority can be held responsible for them.
\end{acknowledgements}

%

\bibliographystyle{aa}
\bibliography{biblio} 

@article{Bellini:2014fua,
    author = "Bellini, Emilio and Sawicki, Ignacy",
    title = "{Maximal freedom at minimum cost: linear large-scale structure in general modifications of gravity}",
    eprint = "1404.3713",
    archivePrefix = "arXiv",
    primaryClass = "astro-ph.CO",
    doi = "10.1088/1475-7516/2014/07/050",
    journal = "JCAP",
    volume = "07",
    pages = "050",
    year = "2014"
}

@article{ICS_emantis,
    author = "S{\'a}ez-Casares, I{\~n}igo and Rasera, Yann and Li, Baojiu",
    title = "{The e-MANTIS emulator: fast predictions of the non-linear matter power spectrum in f(R)CDM cosmology}",
    eprint = "2303.08899",
    archivePrefix = "arXiv",
    primaryClass = "astro-ph.CO",
    doi = "10.1093/mnras/stad3343",
    journal = "Mon. Not. Roy. Astron. Soc.",
    volume = "527",
    number = "3",
    pages = "7242--7262",
    year = "2024"
}

@article{Sefusatti:2015aex,
    author = "Sefusatti, Emiliano and Crocce, Martin and Scoccimarro, Roman and Couchman, Hugh",
    title = "{Accurate Estimators of Correlation Functions in Fourier Space}",
    eprint = "1512.07295",
    archivePrefix = "arXiv",
    primaryClass = "astro-ph.CO",
    doi = "10.1093/mnras/stw1229",
    journal = "Mon. Not. Roy. Astron. Soc.",
    volume = "460",
    number = "4",
    pages = "3624--3636",
    year = "2016"
}

@article{Lu:2025gki,
    author = "Lu, Zhiyu and Simon, Th{\'e}o and Zhang, Pierre",
    title = "{Preference for evolving dark energy in light of the galaxy bispectrum}",
    eprint = "2503.04602",
    archivePrefix = "arXiv",
    primaryClass = "astro-ph.CO",
    month = "3",
    year = "2025"
}

@MISC{Pylians,
    author = {{Villaescusa-Navarro}, Francisco},
    title = "{Pylians: Python libraries for the analysis of numerical simulations}",
    howpublished = {Astrophysics Source Code Library, record ascl:1811.008},
    year = 2018,
    month = nov,
    eid = {ascl:1811.008},
    pages = {ascl:1811.008},
    archivePrefix = {ascl},
    eprint = {1811.008},
    adsurl = {https://ui.adsabs.harvard.edu/abs/2018ascl.soft11008V}
}

@article{cusin,
    author = "Cusin, Giulia and Lewandowski, Matthew and Vernizzi, Filippo",
    title = "{Nonlinear Effective Theory of Dark Energy}",
    eprint = "1712.02782",
    archivePrefix = "arXiv",
    primaryClass = "astro-ph.CO",
    doi = "10.1088/1475-7516/2018/04/061",
    journal = "JCAP",
    volume = "04",
    pages = "061",
    year = "2018"
}

@ARTICLE{martel1998,
   author = {{Martel}, H. and {Shapiro}, P.~R.},
    title = "{A convenient set of comoving cosmological variables and their application}",
  journal = {MNRAS},
   eprint = {astro-ph/9710119},
     year = 1998,
    month = jun,
   volume = 297,
    pages = {467-485},
      doi = {10.1046/j.1365-8711.1998.01497.x},
   adsurl = {http://cdsads.u-strasbg.fr/abs/1998MNRAS.297..467M}
}

@article{llinares,
    author = "Llinares, Claudio",
    title = "{Simulation techniques for modified gravity}",
    eprint = "2103.10890",
    archivePrefix = "arXiv",
    primaryClass = "astro-ph.CO",
    doi = "10.1142/S0218271818480036",
    journal = "Int. J. Mod. Phys. D",
    volume = "27",
    number = "15",
    pages = "1848003",
    year = "2018"
}

@article{Li:2013nua,
    author = "Li, Baojiu and Zhao, Gong-Bo and Koyama, Kazuya",
    title = "{Exploring Vainshtein mechanism on adaptively refined meshes}",
    eprint = "1303.0008",
    archivePrefix = "arXiv",
    primaryClass = "astro-ph.CO",
    doi = "10.1088/1475-7516/2013/05/023",
    journal = "JCAP",
    volume = "05",
    pages = "023",
    year = "2013"
}

@article{Barreira:2013eea,
    author = "Barreira, Alexandre and Li, Baojiu and Hellwing, Wojciech A. and Baugh, Carlton M. and Pascoli, Silvia",
    title = "{Nonlinear structure formation in the Cubic Galileon gravity model}",
    eprint = "1306.3219",
    archivePrefix = "arXiv",
    primaryClass = "astro-ph.CO",
    doi = "10.1088/1475-7516/2013/10/027",
    journal = "JCAP",
    volume = "10",
    pages = "027",
    year = "2013"
}

@article{Becker:2020azq,
    author = "Becker, Christoph and Arnold, Christian and Li, Baojiu and Heisenberg, Lavinia",
    title = "{Proca-stinated cosmology. Part I. A N-body code for the vector Galileon}",
    eprint = "2007.03042",
    archivePrefix = "arXiv",
    primaryClass = "astro-ph.CO",
    doi = "10.1088/1475-7516/2020/10/055",
    journal = "JCAP",
    volume = "10",
    pages = "055",
    year = "2020"
}

@ARTICLE{guillet2011simple,
   author = {{Guillet}, T. and {Teyssier}, R.},
    title = "{A simple multigrid scheme for solving the Poisson equation with arbitrary domain boundaries}",
  journal = {Journal of Computational Physics},
archivePrefix = "arXiv",
   eprint = {1104.1703},
 primaryClass = "physics.comp-ph",
     year = 2011,
    month = jun,
   volume = 230,
    pages = {4756-4771},
      doi = {10.1016/j.jcp.2011.02.044},
   adsurl = {http://cdsads.u-strasbg.fr/abs/2011JCoPh.230.4756G}
}

@article{pysco,
    author = "Breton, Michel-Andr{\`e}s",
    title = "{PySCo: A fast particle-mesh N-body code for modified gravity simulations in Python}",
    eprint = "2410.20501",
    archivePrefix = "arXiv",
    primaryClass = "astro-ph.CO",
    doi = "10.1051/0004-6361/202452770",
    journal = "Astron. Astrophys.",
    volume = "695",
    pages = "A170",
    year = "2025"
}

@ARTICLE{ecosmog,
       author = {{Li}, Baojiu and {Zhao}, Gong-Bo and {Teyssier}, Romain and {Koyama}, Kazuya},
        title = "{ECOSMOG: an Efficient COde for Simulating MOdified Gravity}",
      journal = {\jcap},
         year = 2012,
        month = jan,
       volume = {2012},
       number = {1},
          eid = {051},
        pages = {051},
          doi = {10.1088/1475-7516/2012/01/051},
archivePrefix = {arXiv},
       eprint = {1110.1379},
 primaryClass = {astro-ph.CO},
       adsurl = {https://ui.adsabs.harvard.edu/abs/2012JCAP...01..051L}
}

@ARTICLE{camb,
       author = {{Lewis}, Antony and {Challinor}, Anthony and {Lasenby}, Anthony},
        title = "{Efficient Computation of Cosmic Microwave Background Anisotropies in Closed Friedmann-Robertson-Walker Models}",
      journal = {\apj},
         year = 2000,
        month = aug,
       volume = {538},
       number = {2},
        pages = {473-476},
          doi = {10.1086/309179},
archivePrefix = {arXiv},
       eprint = {astro-ph/9911177},
 primaryClass = {astro-ph},
       adsurl = {https://ui.adsabs.harvard.edu/abs/2000ApJ...538..473L}
}

@article{Hassani:2020rxd,
    author = "Hassani, Farbod and Lombriser, Lucas",
    title = "{$N$-body simulations for parametrized modified gravity}",
    eprint = "2003.05927",
    archivePrefix = "arXiv",
    primaryClass = "astro-ph.CO",
    doi = "10.1093/mnras/staa2083",
    journal = "Mon. Not. Roy. Astron. Soc.",
    volume = "497",
    number = "2",
    pages = "1885--1894",
    year = "2020"
}

@article{Ruan:2021wup,
    author = "Ruan, Cheng-Zong and Hern{\'a}ndez-Aguayo, C{\'e}sar and Li, Baojiu and Arnold, Christian and Baugh, Carlton M. and Klypin, Anatoly and Prada, Francisco",
    title = "{Fast full N-body simulations of generic modified gravity: conformal coupling models}",
    eprint = "2110.00328",
    archivePrefix = "arXiv",
    primaryClass = "astro-ph.CO",
    doi = "10.1088/1475-7516/2022/05/018",
    journal = "JCAP",
    volume = "05",
    number = "05",
    pages = "018",
    year = "2022"
}

@article{Wright:2022krq,
    author = "Wright, Bill S. and Sen Gupta, Ashim and Baker, Tessa and Valogiannis, Georgios and Fiorini, Bartolomeo",
    collaboration = "LSST Dark Energy Science",
    title = "{Hi-COLA: fast, approximate simulations of structure formation in Horndeski gravity}",
    eprint = "2209.01666",
    archivePrefix = "arXiv",
    primaryClass = "astro-ph.CO",
    doi = "10.1088/1475-7516/2023/03/040",
    journal = "JCAP",
    volume = "03",
    pages = "040",
    year = "2023"
}

@ARTICLE{Brando,
       author = {{Brando}, Guilherme and {Koyama}, Kazuya and {Winther}, Hans A.},
        title = "{Revisiting Vainshtein screening for fast N-body simulations}",
      journal = {\jcap},
         year = 2023,
        month = jun,
       volume = {2023},
       number = {6},
          eid = {045},
        pages = {045},
          doi = {10.1088/1475-7516/2023/06/045},
archivePrefix = {arXiv},
       eprint = {2303.09549},
 primaryClass = {astro-ph.CO},
       adsurl = {https://ui.adsabs.harvard.edu/abs/2023JCAP...06..045B}
}

@ARTICLE{Will2014,
       author = {{Will}, Clifford M.},
        title = "{The Confrontation between General Relativity and Experiment}",
      journal = {Living Reviews in Relativity},
         year = 2014,
        month = dec,
       volume = {17},
       number = {1},
          eid = {4},
        pages = {4},
          doi = {10.12942/lrr-2014-4},
archivePrefix = {arXiv},
       eprint = {1403.7377},
 primaryClass = {gr-qc},
       adsurl = {https://ui.adsabs.harvard.edu/abs/2014LRR....17....4W}
}

@ARTICLE{Bellini:hiclass,
       author = {{Bellini}, Emilio and {Sawicki}, Ignacy and {Zumalac{\'a}rregui}, Miguel},
        title = "{hi\_class background evolution, initial conditions and approximation schemes}",
      journal = {\jcap},
         year = 2020,
        month = feb,
       volume = {2020},
       number = {2},
          eid = {008},
        pages = {008},
          doi = {10.1088/1475-7516/2020/02/008},
archivePrefix = {arXiv},
       eprint = {1909.01828},
 primaryClass = {astro-ph.CO},
       adsurl = {https://ui.adsabs.harvard.edu/abs/2020JCAP...02..008B}
}

@article{DESI:2025zgx,
    author = "Abdul Karim, M. and others",
    collaboration = "DESI",
    title = "{DESI DR2 results. II. Measurements of baryon acoustic oscillations and cosmological constraints}",
    eprint = "2503.14738",
    archivePrefix = "arXiv",
    primaryClass = "astro-ph.CO",
    reportNumber = "FERMILAB-PUB-25-0169-PPD",
    doi = "10.1103/tr6y-kpc6",
    journal = "Phys. Rev. D",
    volume = "112",
    number = "8",
    pages = "083515",
    year = "2025"
}

@article{Creminelli:2017sry,
    author = "Creminelli, Paolo and Vernizzi, Filippo",
    title = "{Dark Energy after GW170817 and GRB170817A}",
    eprint = "1710.05877",
    archivePrefix = "arXiv",
    primaryClass = "astro-ph.CO",
    doi = "10.1103/PhysRevLett.119.251302",
    journal = "Phys. Rev. Lett.",
    volume = "119",
    number = "25",
    pages = "251302",
    year = "2017"
}

@article{Ligo_gw170817,
    author = "Abbott, B. P. and others",
    collaboration = "LIGO Scientific, Virgo, Fermi-GBM, INTEGRAL",
    title = "{Gravitational Waves and Gamma-rays from a Binary Neutron Star Merger: GW170817 and GRB 170817A}",
    eprint = "1710.05834",
    archivePrefix = "arXiv",
    primaryClass = "astro-ph.HE",
    reportNumber = "LIGO-P1700308",
    doi = "10.3847/2041-8213/aa920c",
    journal = "Astrophys. J. Lett.",
    volume = "848",
    number = "2",
    pages = "L13",
    year = "2017"
}

@article{DES:2025bxy,
    author = "Abbott, T. M. C. and others",
    collaboration = "DES",
    title = "{Dark Energy Survey: implications for cosmological expansion models from the final DES Baryon Acoustic Oscillation and Supernova data}",
    eprint = "2503.06712",
    archivePrefix = "arXiv",
    primaryClass = "astro-ph.CO",
    reportNumber = "FERMILAB-PUB-25-0127-PPD, DES-2024-0849, FERMILAB-PUB-25-0127-PPD",
    month = "3",
    year = "2025"
}

@INPROCEEDINGS{DESI1,
       author = {{Levi}, Michael and {Allen}, Lori E. and {Raichoor}, Anand and {Baltay}, Charles and {BenZvi}, Segev and {Beutler}, Florian and {Bolton}, Adam and {Castander}, Francisco J. and {Chuang}, Chia-Hsun and {Cooper}, Andrew and {Cuby}, Jean-Gabriel and {Dey}, Arjun and {Eisenstein}, Daniel and {Fan}, Xiaohui and {Flaugher}, Brenna and {Frenk}, Carlos and {Gonzalez-Morales}, Alma X. and {Graur}, Or and {Guy}, Julien and {Habib}, Salman and {Honscheid}, Klaus and {Juneau}, Stephanie and {Kneib}, Jean-Paul and {Lahav}, Ofer and {Lang}, Dustin and {Leauthaud}, Alexie and {Lusso}, Betta and {de la Macorra}, Axel and {Manera}, Marc and {Martini}, Paul and {Mao}, Shude and {Newman}, Jeffrey A. and {Palanque-Delabrouille}, Nathalie and {Percival}, Will J. and {Allende Prieto}, Carlos and {Rockosi}, Constance M. and {Ruhlmann-Kleider}, Vanina and {Schlegel}, David and {Seo}, Hee-Jong and {Song}, Yong-Seon and {Tarle}, Greg and {Wechsler}, Risa and {Weinberg}, David and {Yeche}, Christophe and {Zu}, Ying},
        title = "{The Dark Energy Spectroscopic Instrument (DESI)}",
    booktitle = {Bulletin of the American Astronomical Society},
         year = 2019,
       volume = {51},
        month = sep,
          eid = {57},
        pages = {57},
          doi = {10.48550/arXiv.1907.10688},
archivePrefix = {arXiv},
       eprint = {1907.10688},
 primaryClass = {astro-ph.IM},
       adsurl = {https://ui.adsabs.harvard.edu/abs/2019BAAS...51g..57L}
}

@ARTICLE{LSST2012,
       author = {{LSST Dark Energy Science Collaboration}},
        title = "{Large Synoptic Survey Telescope: Dark Energy Science Collaboration}",
      journal = {arXiv e-prints},
         year = 2012,
        month = nov,
          eid = {arXiv:1211.0310},
        pages = {arXiv:1211.0310},
          doi = {10.48550/arXiv.1211.0310},
archivePrefix = {arXiv},
       eprint = {1211.0310},
 primaryClass = {astro-ph.CO},
       adsurl = {https://ui.adsabs.harvard.edu/abs/2012arXiv1211.0310L}
}

@ARTICLE{Euclid1,
       author = {{Euclid Collaboration} and {Mellier}, Y. and {Abdurro'uf} and {Acevedo Barroso}, J.~A. and {Ach{\'u}carro}, A. and {Adamek}, J. and {Adam}, R. and {Addison}, G.~E. and {Aghanim}, N. and {Aguena}, M. and {Ajani}, V. and {Akrami}, Y. and {Al-Bahlawan}, A. and {Alavi}, A. and {Albuquerque}, I.~S. and {Alestas}, G. and {Alguero}, G. and {Allaoui}, A. and {Allen}, S.~W. and {Allevato}, V. and {Alonso-Tetilla}, A.~V. and {Altieri}, B. and {Alvarez-Candal}, A. and {Alvi}, S. and {Amara}, A. and {Amendola}, L. and {Amiaux}, J. and {Andika}, I.~T. and {Andreon}, S. and {Andrews}, A. and {Angora}, G. and {Angulo}, R.~E. and {Annibali}, F. and {Anselmi}, A. and {Anselmi}, S. and {Arcari}, S. and {Archidiacono}, M. and {Aric{\`o}}, G. and {Arnaud}, M. and {Arnouts}, S. and {Asgari}, M. and {Asorey}, J. and {Atayde}, L. and {Atek}, H. and {Atrio-Barandela}, F. and {Aubert}, M. and {Aubourg}, E. and {Auphan}, T. and {Auricchio}, N. and {Aussel}, B. and {Aussel}, H. and {Avelino}, P.~P. and {Avgoustidis}, A. and {Avila}, S. and {Awan}, S. and {Azzollini}, R. and {Baccigalupi}, C. and {Bachelet}, E. and {Bacon}, D. and {Baes}, M. and {Bagley}, M.~B. and {Bahr-Kalus}, B. and {Balaguera-Antolinez}, A. and {Balbinot}, E. and {Balcells}, M. and {Baldi}, M. and {Baldry}, I. and {Balestra}, A. and {Ballardini}, M. and {Ballester}, O. and {Balogh}, M. and {Ba{\~n}ados}, E. and {Barbier}, R. and {Bardelli}, S. and {Baron}, M. and {Barreiro}, T. and {Barrena}, R. and {Barriere}, J.-C. and {Barros}, B.~J. and {Barthelemy}, A. and {Bartolo}, N. and {Basset}, A. and {Battaglia}, P. and {Battisti}, A.~J. and {Baugh}, C.~M. and {Baumont}, L. and {Bazzanini}, L. and {Beaulieu}, J.-P. and {Beckmann}, V. and {Belikov}, A.~N. and {Bel}, J. and {Bellagamba}, F. and {Bella}, M. and {Bellini}, E. and {Benabed}, K. and {Bender}, R. and {Benevento}, G. and {Bennett}, C.~L. and {Benson}, K. and {Bergamini}, P. and {Bermejo-Climent}, J.~R. and {Bernardeau}, F. and {Bertacca}, D. and {Berthe}, M. and {Berthier}, J. and {Bethermin}, M. and {Beutler}, F. and {Bevillon}, C. and {Bhargava}, S. and {Bhatawdekar}, R. and {Bianchi}, D. and {Bisigello}, L. and {Biviano}, A. and {Blake}, R.~P. and {Blanchard}, A. and {Blazek}, J. and {Blot}, L. and {Bosco}, A. and {Bodendorf}, C. and {Boenke}, T. and {B{\"o}hringer}, H. and {Boldrini}, P. and {Bolzonella}, M. and {Bonchi}, A. and {Bonici}, M. and {Bonino}, D. and {Bonino}, L. and {Bonvin}, C. and {Bon}, W. and {Booth}, J.~T. and {Borgani}, S. and {Borlaff}, A.~S. and {Borsato}, E. and {Bose}, B. and {Botticella}, M.~T. and {Boucaud}, A. and {Bouche}, F. and {Boucher}, J.~S. and {Boutigny}, D. and {Bouvard}, T. and {Bouwens}, R. and {Bouy}, H. and {Bowler}, R.~A.~A. and {Bozza}, V. and {Bozzo}, E. and {Branchini}, E. and {Brando}, G. and {Brau-Nogue}, S. and {Brekke}, P. and {Bremer}, M.~N. and {Brescia}, M. and {Breton}, M.-A. and {Brinchmann}, J. and {Brinckmann}, T. and {Brockley-Blatt}, C. and {Brodwin}, M. and {Brouard}, L. and {Brown}, M.~L. and {Bruton}, S. and {Bucko}, J. and {Buddelmeijer}, H. and {Buenadicha}, G. and {Buitrago}, F. and {Burger}, P. and {Burigana}, C. and {Busillo}, V. and {Busonero}, D. and {Cabanac}, R. and {Cabayol-Garcia}, L. and {Cagliari}, M.~S. and {Caillat}, A. and {Caillat}, L. and {Calabrese}, M. and {Calabro}, A. and {Calderone}, G. and {Calura}, F. and {Camacho Quevedo}, B. and {Camera}, S. and {Campos}, L. and {Ca{\~n}as-Herrera}, G. and {Candini}, G.~P. and {Cantiello}, M. and {Capobianco}, V. and {Cappellaro}, E. and {Cappelluti}, N. and {Cappi}, A. and {Caputi}, K.~I. and {Cara}, C. and {Carbone}, C. and {Cardone}, V.~F. and {Carella}, E. and {Carlberg}, R.~G. and {Carle}, M. and {Carminati}, L. and {Caro}, F. and {Carrasco}, J.~M. and {Carretero}, J. and {Carrilho}, P. and {Carron Duque}, J. and {Carry}, B.},
        title = "{Euclid: I. Overview of the Euclid mission}",
      journal = {\aap},
         year = 2025,
        month = may,
       volume = {697},
          eid = {A1},
        pages = {A1},
          doi = {10.1051/0004-6361/202450810},
archivePrefix = {arXiv},
       eprint = {2405.13491},
 primaryClass = {astro-ph.CO},
       adsurl = {https://ui.adsabs.harvard.edu/abs/2025A&A...697A...1E}
}

@article{Winther:2015wla,
    author = "Winther, Hans A. and others",
    title = "{Modified Gravity N-body Code Comparison Project}",
    eprint = "1506.06384",
    archivePrefix = "arXiv",
    primaryClass = "astro-ph.CO",
    doi = "10.1093/mnras/stv2253",
    journal = "Mon. Not. Roy. Astron. Soc.",
    volume = "454",
    number = "4",
    pages = "4208--4234",
    year = "2015"
}

@article{Planck:2018vyg,
    author = "Aghanim, N. and others",
    collaboration = "Planck",
    title = "{Planck 2018 results. VI. Cosmological parameters}",
    eprint = "1807.06209",
    archivePrefix = "arXiv",
    primaryClass = "astro-ph.CO",
    doi = "10.1051/0004-6361/201833910",
    journal = "Astron. Astrophys.",
    volume = "641",
    pages = "A6",
    year = "2020",
    note = "[Erratum: Astron.Astrophys. 652, C4 (2021)]"
}

@ARTICLE{Winther_QSA,
       author = {{Winther}, Hans A. and {Ferreira}, Pedro G.},
        title = "{Vainshtein mechanism beyond the quasistatic approximation}",
      journal = {\prd},
         year = 2015,
        month = sep,
       volume = {92},
       number = {6},
          eid = {064005},
        pages = {064005},
          doi = {10.1103/PhysRevD.92.064005},
archivePrefix = {arXiv},
       eprint = {1505.03539},
 primaryClass = {gr-qc},
       adsurl = {https://ui.adsabs.harvard.edu/abs/2015PhRvD..92f4005W}
}

@article{Moretti:2026axy,
    author = "Moretti, Tommaso and Frusciante, Noemi and Verza, Giovanni and Pace, Francesco",
    title = "{How deep can a cosmic void be? Voids-informed theoretical bounds in Galileon gravity}",
    eprint = "2601.05145",
    archivePrefix = "arXiv",
    primaryClass = "astro-ph.CO",
    month = "1",
    year = "2026"
}

@ARTICLE{mpgrafic,
       author = {{Prunet}, S. and {Pichon}, C. and {Aubert}, D. and {Pogosyan}, D. and {Teyssier}, R. and {Gottloeber}, S.},
        title = "{Initial Conditions For Large Cosmological Simulations}",
      journal = {\apjs},
         year = 2008,
        month = oct,
       volume = {178},
       number = {2},
        pages = {179-188},
          doi = {10.1086/590370},
archivePrefix = {arXiv},
       eprint = {0804.3536},
 primaryClass = {astro-ph},
       adsurl = {https://ui.adsabs.harvard.edu/abs/2008ApJS..178..179P}
}

@ARTICLE{Rasera22_RayGal,
       author = {{Rasera}, Y. and {Breton}, M.-A. and {Corasaniti}, P.-S. and {Allingham}, J. and {Roy}, F. and {Reverdy}, V. and {Pellegrin}, T. and {Saga}, S. and {Taruya}, A. and {Agarwal}, S. and {Anselmi}, S.},
        title = "{The RayGalGroupSims cosmological simulation suite for the study of relativistic effects: An application to lensing-matter clustering statistics}",
      journal = {\aap},
         year = 2022,
        month = may,
       volume = {661},
          eid = {A90},
        pages = {A90},
          doi = {10.1051/0004-6361/202141908},
archivePrefix = {arXiv},
       eprint = {2111.08745},
 primaryClass = {astro-ph.CO},
       adsurl = {https://ui.adsabs.harvard.edu/abs/2022A&A...661A..90R}
}

@ARTICLE{Gubitosi13_EFTofDE,
       author = {{Gubitosi}, Giulia and {Piazza}, Federico and {Vernizzi}, Filippo},
        title = "{The effective field theory of dark energy}",
      journal = {\jcap},
         year = 2013,
        month = feb,
       volume = {2013},
       number = {2},
          eid = {032},
        pages = {032},
          doi = {10.1088/1475-7516/2013/02/032},
archivePrefix = {arXiv},
       eprint = {1210.0201},
 primaryClass = {hep-th},
       adsurl = {https://ui.adsabs.harvard.edu/abs/2013JCAP...02..032G}
}

@ARTICLE{Frusciante20_EFTofDE_review,
       author = {{Frusciante}, Noemi and {Perenon}, Louis},
        title = "{Effective field theory of dark energy: A review}",
      journal = {\physrep},
         year = 2020,
        month = may,
       volume = {857},
        pages = {1-63},
          doi = {10.1016/j.physrep.2020.02.004},
archivePrefix = {arXiv},
       eprint = {1907.03150},
 primaryClass = {astro-ph.CO},
       adsurl = {https://ui.adsabs.harvard.edu/abs/2020PhR...857....1F}
}

@ARTICLE{Hu14_EFTCAMB,
       author = {{Hu}, Bin and {Raveri}, Marco and {Frusciante}, Noemi and {Silvestri}, Alessandra},
        title = "{Effective field theory of cosmic acceleration: An implementation in CAMB}",
      journal = {\prd},
         year = 2014,
        month = may,
       volume = {89},
       number = {10},
          eid = {103530},
        pages = {103530},
          doi = {10.1103/PhysRevD.89.103530},
archivePrefix = {arXiv},
       eprint = {1312.5742},
 primaryClass = {astro-ph.CO},
       adsurl = {https://ui.adsabs.harvard.edu/abs/2014PhRvD..89j3530H}
}

@article{Breton:2018wzk,
    author = "Breton, Michel-Andr{\`e}s and Rasera, Yann and Taruya, Atsushi and Lacombe, Osmin and Saga, Shohei",
    title = "{Imprints of relativistic effects on the asymmetry of the halo cross-correlation function: from linear to non-linear scales}",
    eprint = "1803.04294",
    archivePrefix = "arXiv",
    primaryClass = "astro-ph.CO",
    reportNumber = "YITP-18-16",
    doi = "10.1093/mnras/sty3206",
    journal = "Mon. Not. Roy. Astron. Soc.",
    volume = "483",
    number = "2",
    pages = "2671--2696",
    year = "2019"
}

@ARTICLE{Frusciante17_nonlinearEFTofDE,
       author = {{Frusciante}, Noemi and {Papadomanolakis}, Georgios},
        title = "{Tackling non-linearities with the effective field theory of dark energy and modified gravity}",
      journal = {\jcap},
         year = 2017,
        month = dec,
       volume = {2017},
       number = {12},
          eid = {014},
        pages = {014},
          doi = {10.1088/1475-7516/2017/12/014},
archivePrefix = {arXiv},
       eprint = {1706.02719},
 primaryClass = {gr-qc},
       adsurl = {https://ui.adsabs.harvard.edu/abs/2017JCAP...12..014F}
}

@ARTICLE{Bellini15_bispectrum,
       author = {{Bellini}, Emilio and {Jimenez}, Raul and {Verde}, Licia},
        title = "{Signatures of Horndeski gravity on the dark matter bispectrum}",
      journal = {\jcap},
         year = 2015,
        month = may,
       volume = {2015},
       number = {5},
        pages = {057-057},
          doi = {10.1088/1475-7516/2015/05/057},
archivePrefix = {arXiv},
       eprint = {1504.04341},
 primaryClass = {astro-ph.CO},
       adsurl = {https://ui.adsabs.harvard.edu/abs/2015JCAP...05..057B}
}

@ARTICLE{Yamauchi17_bispectrum,
       author = {{Yamauchi}, Daisuke and {Yokoyama}, Shuichiro and {Tashiro}, Hiroyuki},
        title = "{Constraining modified theories of gravity with the galaxy bispectrum}",
      journal = {\prd},
         year = 2017,
        month = dec,
       volume = {96},
       number = {12},
          eid = {123516},
        pages = {123516},
          doi = {10.1103/PhysRevD.96.123516},
archivePrefix = {arXiv},
       eprint = {1709.03243},
 primaryClass = {astro-ph.CO},
       adsurl = {https://ui.adsabs.harvard.edu/abs/2017PhRvD..96l3516Y}
}

@ARTICLE{Cusin18_EFTofLSS,
       author = {{Cusin}, Giulia and {Lewandowski}, Matthew and {Vernizzi}, Filippo},
        title = "{Dark energy and modified gravity in the Effective Field Theory of Large-Scale Structure}",
      journal = {\jcap},
         year = 2018,
        month = apr,
       volume = {2018},
       number = {4},
          eid = {005},
        pages = {005},
          doi = {10.1088/1475-7516/2018/04/005},
archivePrefix = {arXiv},
       eprint = {1712.02783},
 primaryClass = {astro-ph.CO},
       adsurl = {https://ui.adsabs.harvard.edu/abs/2018JCAP...04..005C}
}

@ARTICLE{Bose23_halomodel_EFTofDE,
       author = {{Bose}, B. and {Tsedrik}, M. and {Kennedy}, J. and {Lombriser}, L. and {Pourtsidou}, A. and {Taylor}, A.},
        title = "{Fast and accurate predictions of the non-linear matter power spectrum for general models of Dark Energy and Modified Gravity}",
      journal = {\mnras},
         year = 2023,
        month = mar,
       volume = {519},
       number = {3},
        pages = {4780-4800},
          doi = {10.1093/mnras/stac3783},
archivePrefix = {arXiv},
       eprint = {2210.01094},
 primaryClass = {astro-ph.CO},
       adsurl = {https://ui.adsabs.harvard.edu/abs/2023MNRAS.519.4780B}
}

@ARTICLE{deBoe_halomodel_EFTofDE,
       author = {{de Boe}, Dani and {Pantiri}, Mattia and {Ye}, Gen and {Silvestri}, Alessandra},
        title = "{Nonlinear Scales in Luminal Horndeski -- I. Halo mass function and power spectrum boost in models with Vainshtein screening}",
      journal = {arXiv e-prints},
         year = 2026,
        month = jan,
          eid = {arXiv:2601.02074},
        pages = {arXiv:2601.02074},
          doi = {10.48550/arXiv.2601.02074},
archivePrefix = {arXiv},
       eprint = {2601.02074},
 primaryClass = {astro-ph.CO},
       adsurl = {https://ui.adsabs.harvard.edu/abs/2026arXiv260102074D}
}

@article{Bloomfield:2012ff,
    author = "Bloomfield, Jolyon K. and Flanagan, {\'E}anna {\'E}. and Park, Minjoon and Watson, Scott",
    title = "{Dark energy or modified gravity?  An effective field theory approach}",
    eprint = "1211.7054",
    archivePrefix = "arXiv",
    primaryClass = "astro-ph.CO",
    doi = "10.1088/1475-7516/2013/08/010",
    journal = "JCAP",
    volume = "08",
    pages = "010",
    year = "2013"
}

@article{Zumalacarregui:2016pph,
    author = "Zumalac{\'a}rregui, Miguel and Bellini, Emilio and Sawicki, Ignacy and Lesgourgues, Julien and Ferreira, Pedro G.",
    title = "{hi{\_}class: Horndeski in the Cosmic Linear Anisotropy Solving System}",
    eprint = "1605.06102",
    archivePrefix = "arXiv",
    primaryClass = "astro-ph.CO",
    reportNumber = "NORDITA-2016-41",
    doi = "10.1088/1475-7516/2017/08/019",
    journal = "JCAP",
    volume = "08",
    pages = "019",
    year = "2017"
}

@article{Bellini:2017avd,
    author = "Bellini, E. and others",
    title = "{Comparison of Einstein-Boltzmann solvers for testing general relativity}",
    eprint = "1709.09135",
    archivePrefix = "arXiv",
    primaryClass = "astro-ph.CO",
    reportNumber = "NORDITA-2017-098",
    doi = "10.1103/PhysRevD.97.023520",
    journal = "Phys. Rev. D",
    volume = "97",
    number = "2",
    pages = "023520",
    year = "2018"
}

@article{Nouri-Zonoz:2025cul,
    author = "Nouri-Zonoz, Ahmad and Hassani, Farbod and Bellini, Emilio and Kunz, Martin",
    title = "{KGB-evolution: a relativistic $N$-body code for kinetic gravity braiding models}",
    eprint = "2511.04676",
    archivePrefix = "arXiv",
    primaryClass = "astro-ph.CO",
    month = "11",
    year = "2025"
}

@article{Hassani:2019lmy,
    author = "Hassani, Farbod and Adamek, Julian and Kunz, Martin and Vernizzi, Filippo",
    title = "{$k$-evolution: a relativistic N-body code for clustering dark energy}",
    eprint = "1910.01104",
    archivePrefix = "arXiv",
    primaryClass = "astro-ph.CO",
    doi = "10.1088/1475-7516/2019/12/011",
    journal = "JCAP",
    volume = "12",
    pages = "011",
    year = "2019"
}

@article{Horndeski:1974wa,
    author = "Horndeski, Gregory Walter",
    title = "{Second-order scalar-tensor field equations in a four-dimensional space}",
    doi = "10.1007/BF01807638",
    journal = "Int. J. Theor. Phys.",
    volume = "10",
    pages = "363--384",
    year = "1974"
}

@article{Deffayet:2011gz,
    author = "Deffayet, C. and Gao, Xian and Steer, D. A. and Zahariade, G.",
    title = "{From k-essence to generalised Galileons}",
    eprint = "1103.3260",
    archivePrefix = "arXiv",
    primaryClass = "hep-th",
    doi = "10.1103/PhysRevD.84.064039",
    journal = "Phys. Rev. D",
    volume = "84",
    pages = "064039",
    year = "2011"
}

@article{Zumalacarregui:2013pma,
    author = "Zumalac{\'a}rregui, Miguel and Garc{\'\i}a-Bellido, Juan",
    title = "{Transforming gravity: from derivative couplings to matter to second-order scalar-tensor theories beyond the Horndeski Lagrangian}",
    eprint = "1308.4685",
    archivePrefix = "arXiv",
    primaryClass = "gr-qc",
    reportNumber = "IFT-UAM-CSIC-13-090",
    doi = "10.1103/PhysRevD.89.064046",
    journal = "Phys. Rev. D",
    volume = "89",
    pages = "064046",
    year = "2014"
}

@article{Gleyzes:2014qga,
    author = "Gleyzes, J{\'e}r{\^o}me and Langlois, David and Piazza, Federico and Vernizzi, Filippo",
    title = "{Exploring gravitational theories beyond Horndeski}",
    eprint = "1408.1952",
    archivePrefix = "arXiv",
    primaryClass = "astro-ph.CO",
    doi = "10.1088/1475-7516/2015/02/018",
    journal = "JCAP",
    volume = "02",
    pages = "018",
    year = "2015"
}

@article{Gleyzes:2014rba,
    author = "Gleyzes, J{\'e}r{\^o}me and Langlois, David and Vernizzi, Filippo",
    title = "{A unifying description of dark energy}",
    eprint = "1411.3712",
    archivePrefix = "arXiv",
    primaryClass = "hep-th",
    doi = "10.1142/S021827181443010X",
    journal = "Int. J. Mod. Phys. D",
    volume = "23",
    number = "13",
    pages = "1443010",
    year = "2015"
}

@article{Langlois:2015cwa,
    author = "Langlois, David and Noui, Karim",
    title = "{Degenerate higher derivative theories beyond Horndeski: evading the Ostrogradski instability}",
    eprint = "1510.06930",
    archivePrefix = "arXiv",
    primaryClass = "gr-qc",
    doi = "10.1088/1475-7516/2016/02/034",
    journal = "JCAP",
    volume = "02",
    pages = "034",
    year = "2016"
}

@article{Langlois:2018dxi,
    author = "Langlois, David",
    title = "{Dark energy and modified gravity in degenerate higher-order scalar{\textendash}tensor (DHOST) theories: A review}",
    eprint = "1811.06271",
    archivePrefix = "arXiv",
    primaryClass = "gr-qc",
    doi = "10.1142/S0218271819420069",
    journal = "Int. J. Mod. Phys. D",
    volume = "28",
    number = "05",
    pages = "1942006",
    year = "2019"
}

@article{Langlois:2017mxy,
    author = "Langlois, David and Mancarella, Michele and Noui, Karim and Vernizzi, Filippo",
    title = "{Effective Description of Higher-Order Scalar-Tensor Theories}",
    eprint = "1703.03797",
    archivePrefix = "arXiv",
    primaryClass = "hep-th",
    doi = "10.1088/1475-7516/2017/05/033",
    journal = "JCAP",
    volume = "05",
    pages = "033",
    year = "2017"
}

@article{Gupta:2024seu,
    author = "Gupta, Ashim Sen and Fiorini, Bartolomeo and Baker, Tessa",
    title = "{K-mouflage at high k: extending the reach of Hi-COLA}",
    eprint = "2407.00855",
    archivePrefix = "arXiv",
    primaryClass = "astro-ph.CO",
    doi = "10.1088/1475-7516/2024/11/052",
    journal = "JCAP",
    volume = "11",
    pages = "052",
    year = "2024"
}

@article{Baker:2017hug,
    author = "Baker, T. and Bellini, E. and Ferreira, P. G. and Lagos, M. and Noller, J. and Sawicki, I.",
    title = "{Strong constraints on cosmological gravity from GW170817 and GRB 170817A}",
    eprint = "1710.06394",
    archivePrefix = "arXiv",
    primaryClass = "astro-ph.CO",
    doi = "10.1103/PhysRevLett.119.251301",
    journal = "Phys. Rev. Lett.",
    volume = "119",
    number = "25",
    pages = "251301",
    year = "2017"
}

@article{Sakstein:2017xjx,
    author = "Sakstein, Jeremy and Jain, Bhuvnesh",
    title = "{Implications of the Neutron Star Merger GW170817 for Cosmological Scalar-Tensor Theories}",
    eprint = "1710.05893",
    archivePrefix = "arXiv",
    primaryClass = "astro-ph.CO",
    doi = "10.1103/PhysRevLett.119.251303",
    journal = "Phys. Rev. Lett.",
    volume = "119",
    number = "25",
    pages = "251303",
    year = "2017"
}

@article{Ezquiaga:2017ekz,
    author = "Ezquiaga, Jose Mar{\'\i}a and Zumalac{\'a}rregui, Miguel",
    title = "{Dark Energy After GW170817: Dead Ends and the Road Ahead}",
    eprint = "1710.05901",
    archivePrefix = "arXiv",
    primaryClass = "astro-ph.CO",
    reportNumber = "IFT-UAM-CSIC-17-096, NORDITA-2017-109",
    doi = "10.1103/PhysRevLett.119.251304",
    journal = "Phys. Rev. Lett.",
    volume = "119",
    number = "25",
    pages = "251304",
    year = "2017"
}

@ARTICLE{alimi10,
       author = {{Alimi}, J.-M. and {F{\"u}zfa}, A. and {Boucher}, V. and {Rasera}, Y. and {Courtin}, J. and {Corasaniti}, P.-S.},
        title = "{Imprints of dark energy on cosmic structure formation - I. Realistic quintessence models and the non-linear matter power spectrum}",
      journal = {\mnras},
         year = 2010,
        month = jan,
       volume = {401},
       number = {2},
        pages = {775-790},
          doi = {10.1111/j.1365-2966.2009.15712.x},
archivePrefix = {arXiv},
       eprint = {0903.5490},
 primaryClass = {astro-ph.CO},
       adsurl = {https://ui.adsabs.harvard.edu/abs/2010MNRAS.401..775A}
}

@ARTICLE{colombi09,
       author = {{Colombi}, St{\'e}phane and {Jaffe}, Andrew and {Novikov}, Dmitri and {Pichon}, Christophe},
        title = "{Accurate estimators of power spectra in N-body simulations}",
      journal = {\mnras},
         year = 2009,
        month = feb,
       volume = {393},
       number = {2},
        pages = {511-526},
          doi = {10.1111/j.1365-2966.2008.14176.x},
archivePrefix = {arXiv},
       eprint = {0811.0313},
 primaryClass = {astro-ph},
       adsurl = {https://ui.adsabs.harvard.edu/abs/2009MNRAS.393..511C}
}

@ARTICLE{teyssier02,
       author = {{Teyssier}, R.},
        title = "{Cosmological hydrodynamics with adaptive mesh refinement. A new high resolution code called RAMSES}",
      journal = {\aap},
         year = 2002,
        month = apr,
       volume = {385},
        pages = {337-364},
          doi = {10.1051/0004-6361:20011817},
archivePrefix = {arXiv},
       eprint = {astro-ph/0111367},
 primaryClass = {astro-ph},
       adsurl = {https://ui.adsabs.harvard.edu/abs/2002A&A...385..337T}
}

@ARTICLE{rasera14,
       author = {{Rasera}, Y. and {Corasaniti}, P.-S. and {Alimi}, J.-M. and {Bouillot}, V. and {Reverdy}, V. and {Balm{\`e}s}, I.},
        title = "{Cosmic-variance limited Baryon Acoustic Oscillations from the DEUS-FUR {\ensuremath{\Lambda}}CDM simulation}",
      journal = {\mnras},
         year = 2014,
        month = may,
       volume = {440},
       number = {2},
        pages = {1420-1434},
          doi = {10.1093/mnras/stu295},
archivePrefix = {arXiv},
       eprint = {1311.5662},
 primaryClass = {astro-ph.CO},
       adsurl = {https://ui.adsabs.harvard.edu/abs/2014MNRAS.440.1420R}
}

\begin{appendix}

\section{Convergence tests simulation suite}
\label{sec:convtests_methods}

The \texttt{ECOSMOG-EFT} code, with its ability to run $N$-body simulations with adaptive mesh refinements, is essential to generating the high-resolution density maps of EFTofDE cosmologies. As a result, the code needs to be robust to changes in various simulation parameters. In this section, we describe the methods for an array of convergence tests for \texttt{ECOSMOG-EFT}. We leave out the convergence testing for \texttt{PySCo-EFT}; as demonstrated in Sec.~\ref{sec:validation_results}, the results from the two codes show excellent agreement when PM-only simulations are used for comparison. 

We use the ``$\Lambda$CDM'' version of the background cosmology from the RayGal Simulation Suite \citep{Breton:2018wzk,Rasera22_RayGal} for all tests, which is specified in Table~\ref{tab:conv_params}. We start all the simulations at $z_{\rm ini} = 56.88$ (unless otherwise specified) and run a pair of simulations (EFTofDE and $\Lambda$CDM) for each run, with 6 refinement levels for both, with a mass refinement threshold of $m_{\rm ref} = 14$ times the particle mass (unless otherwise specified). Between each pair of simulation runs, we use the same initial conditions, generated using 2LPT with our modified \texttt{MPGRAFIC} code and a \texttt{CAMB}-based linear power spectrum. 

For the power spectra in the results, we use the same setup as described in Sec.~\ref{sec:eft_methods}. For the mass resolution test and the last four convergence tests described below, we used a Fourier-space grid with $2048^3$ cells, yielding a $k_{\rm Nyquist}/2 \approx 9.8 h^{-1} \, \rm{Mpc}$. For the box size test, we used a Fourier grid $8$ times finer than the particle grid. 

We conducted convergence tests with respect to six numerical parameters: mass resolution, box size, multigrid smoothing cycles, mass refinement threshold, starting redshift, and scalar field solver convergence criterion. The variations in these parameters are listed in the bottom panel of Table~\ref{tab:conv_params}. For the mass resolution convergence study, the reference parameters are $512^3$ particles and $L_{\rm box} = 328.125\,h^{-1}\,\textrm{Mpc}$. For the box size study, they are $512^3$ particles and $L_{\rm box} = 656.25\,h^{-1}\,\textrm{Mpc}$. For the other four convergence studies, the reference parameters are $256^3$ particles and $L_{\rm box} = 328.125\,h^{-1}\,\textrm{Mpc}$. The results are presented in Appendix~\ref{sec:convtests_results}.

\begin{table}[h]
\centering
\caption{Convergence tests: cosmological and numerical parameters}
\label{tab:conv_params}
\begin{tabular}{lll}
\hline\hline
\textbf{Cosmological parameter} & \textbf{Symbol} & \textbf{Value} \\
\hline\hline
Hubble parameter              & $h$                    & \phantom{$-$}0.720 \\
Matter density                & $\Omega_{\rm m0}$      & \phantom{$-$}0.257 \\
DE density                    & $\Omega_{\Lambda}$     & \phantom{$-$}0.743 \\
\makecell[l]{Power spectrum \\ normalization}
                              & $\sigma_8^{\rm \Lambda CDM}$             & \phantom{$-$}0.801 \\
Scalar spectral index         & $n_{\rm s}$            & \phantom{$-$}0.963 \\
Braiding parameter            & $\alpha_{\text{B0}}$   & $-0.480$           \\
Planck mass run rate          & $\alpha_{\text{M0}}$   & \phantom{$-$}0.000 \\
\hline\hline
\textbf{Numerical parameter} & \textbf{Values} & \textbf{Figure} \\
\hline\hline
Mass resolution
  & $\{64^3, 128^3, 256^3, 512^3\}$
  & \ref{fig:conv_massres} \\
Box size $L_{\rm box}$ $[h^{-1}\,\mathrm{Mpc}]$
  & $\{656.25, 328.125\}$
  & \ref{fig:conv_boxsize} \\
Pre/post smoothing cycles
  & \makecell[l]{$\{1$-$1, 3, 3, 6$-$3,$ \\ $6$-$6, 12$-$3, 12$-$6\}$}
  & \ref{fig:conv_nprenpost} \\
Refinement threshold $m_{\rm ref}$
  & $\{20, 14, 8\}$
  & \ref{fig:conv_mref} \\
Initial redshift $z_{\rm ini}$
  & $\{56.88, 28.41\}$
  & \ref{fig:conv_zstart} \\
Scalar field convergence
  & \makecell[l]{Four \\ multiplicative \\ factors}
  & \ref{fig:conv_chiconv} \\
\hline\hline
\end{tabular}
\end{table}

\paragraph{Mass resolution}

For the first test, we varied the mass resolution of the simulations. We first generated initial conditions of $L_{\rm box} = 328.125\,h^{-1}\,\textrm{Mpc}$ and $512^3$ particles. Subsequently, the density field in these ICs was degraded by averaging; the particle displacements and velocities were generated again for the lower resolutions using second-order Lagrangian perturbation theory. This process was conducted thrice, creating ICs with with $256^3$, $128^3$, and $64^3$ particles with the same $L_{\rm box} = 328.125\,h^{-1}\,\textrm{Mpc}$. We then compared the boosts obtained from the four paired simulation runs.

\paragraph{Box size and variance}

For the next test, we generated initial conditions with $L_{\rm box} = 656.25\,h^{-1}\,\textrm{Mpc}$ and $512^3$ particles and ran an EFT-$\Lambda$CDM pair of simulations with these ICs. The boost $P_{\rm EFT} / P_{\Lambda \rm{CDM}}$ from this pair was computed. Next, the white noise from these ICs was split into eight cubes to generate eight sets of initial conditions of box size $328.125\,h^{-1}\,\textrm{Mpc}$ and $256^3$ particles each. We compared the boost from the $512^3$ particle runs to the mean of the boosts obtained from the eight $256^3$ runs.

\paragraph{Multigrid smoothing cycles}

The multigrid solver for the additional scalar field $\chi$ works by first smoothing an initial guess for $\chi$ with $N_{\rm pre}$ Gauss-Seidel iterations, running the multigrid algorithm, and then smoothing the solution obtained with $N_{\rm post}$ iterations. For this test, we varied $N_{\rm pre}$ and $N_{\rm post}$ between six combinations to assess their impact on the power spectrum boost. For each $N_{\rm pre}\mbox{--}N_{\rm post}$ combination, the simulation pairs were run with 6 refinement levels and $m_{\rm ref} = 14$, in a box of size $328.125 \, \textrm{Mpc}/h$ with $256^3$ particles.

\paragraph{Mass refinement threshold}

The next test varied the mass refinement threshold $m_{\rm ref}$, which is the minimum number of particles in a cell to trigger a finer refinement level. For this test, we used the same ICs as the multigrid smoothing cycles test above. The $m_{\rm ref}$ was varied between the three values $\{20,14,8\}$ and the paired EFT-$\Lambda$CDM simulations were conducted for each $m_{\rm ref}$ with $N_{\rm pre} = N_{\rm post} = 6$. We also conducted a pair of particle-mesh only simulations ($m_{\rm ref} = \infty$) for comparison.

\paragraph{Starting redshift}

For this test, we checked the impact of different starting times for the simulation. For the generation of ICs with \texttt{MPGRAFIC}, we changed the $\sigma_{\rm grid} (z_{\rm ini})$, the RMS fluctuation of the linear power spectrum at the grid scale of the simulation at the starting redshift.  One set of initial conditions was generated at $z_{\rm ini} = 56.88$, corresponding to $\sigma_{\rm grid} (z_{\rm ini}) = 0.05$. Another set of ICs was then computed at $z_{\rm ini} = 28.41$ by setting $\sigma_{\rm grid} (z_{\rm ini}) = 0.1$. With each of these ICs, we ran a pair of EFT-$\Lambda$CDM simulations in a $328.125\,h^{-1}\,\textrm{Mpc}$ box with $256^3$ particles and 6 refinement levels and $(N_{\rm pre}, N_{\rm post}) = (6,6)$ and $m_{\rm ref} = 14$.

\paragraph{Scalar field solver threshold}

The multigrid-Gauss--Seidel solver that computes the additional scalar field $\chi$ terminates based on a combination of stopping criteria. The main condition stops the solver if the residual $d_h < 10^{-9}$ in the code units for the field at the coarsest level and $d_h < 10^{-10}$ at the refinement levels. For the final convergence test, we modified these thresholds by multiplying and dividing them by factors of 10 and 2. We then ran four simulation pairs in boxes of side $328.125\,h^{-1}\,\textrm{Mpc}$, each with $256^3$ particles and 6 refinement levels and $(N_{\rm pre}, N_{\rm post}) = (6,6)$ and $m_{\rm ref} = 14$.

\section{Convergence test results}
\label{sec:convtests_results}

We performed convergence tests for the \texttt{ECOSMOG-EFT} code with respect to six numerical parameters with the setup and cosmology defined in Appendix~\ref{sec:convtests_methods}. The results are described in this section. All power spectra and boosts presented here are at $z=0$.

\paragraph{Mass resolution}

\begin{figure}[h!]
   \centering
   \includegraphics[width=\hsize]{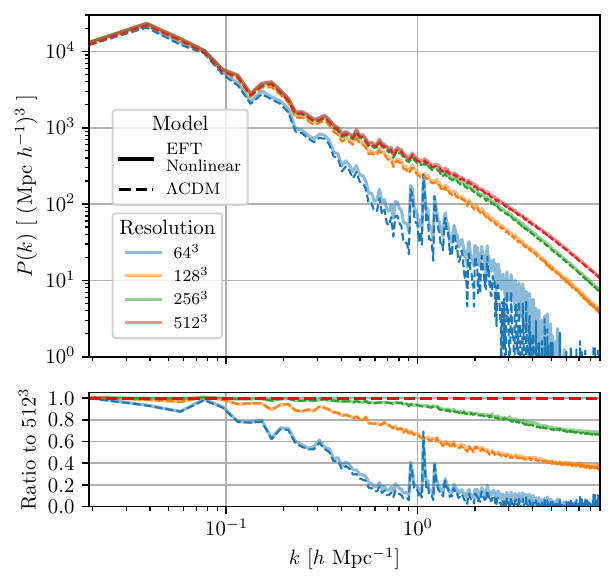}
      \caption{\textit{Top}: The EFT-non-linear (solid) and $\Lambda$CDM (dashed) power spectra from simulations of mass resolution $64^3$, $128^3$, $256^3$ to a $512^3$ simulation. \textit{Bottom}: The ratio of the power spectra to their corresponding $512^3$ version.}
         \label{fig:convt-mres-ps}
   \end{figure}

Fig.~\ref{fig:convt-mres-ps} shows the power spectra in the full non-linear EFT (solid) and the $\Lambda$CDM (dashed) cases for four mass resolution levels, with $64^3$, $128^3$, $256^3$, and $512^3$ particles. The bottom panel plots the ratio of the power spectra for the first three resolution levels to the $512^3$ version for both the EFT and $\Lambda$CDM models. The ratios demonstrate how the power spectra are underestimated at small scales ($k \gtrsim 0.1 \, h \, \rm{Mpc}^{-1}$) for the lower resolution runs. Upon improving the mass resolution, the power spectrum at a given wavenumber converges smoothly towards the power spectrum of the highest resolution run. This mass resolution effect is common to AMR-based simulations and has been reported in earlier \texttt{RAMSES}-based simulations \citep{rasera14,Rasera22_RayGal,ICS_emantis}. 

\begin{figure}[t]
    \centering
    \includegraphics[width=\hsize]{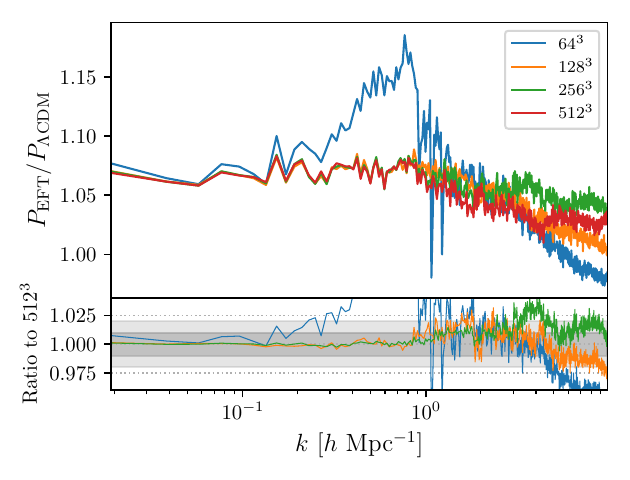}
    \caption{\textbf{Mass Resolution} \textit{Top}: Power spectrum ratios at $z=0$ from four simulation boxes of comoving boxsize $328.125\,h^{-1}\,\rm Mpc$ with different mass resolutions. \textit{Bottom}: Ratios of the curves to the $512^3$ particle simulation.}
    \label{fig:conv_massres}
\end{figure}

It is also shown in \citet{ICS_emantis} that the impact of mass resolution on the boost $R(k)$ is an order of magnitude lower than on the power spectra. The boost was computed for each of the four simulation boxes ($64^3$, $128^3$, $256^3$, $512^3$); it is displayed in the top panel of Fig.~\ref{fig:conv_massres}. The bottom panel shows the ratio of the $64^3$, $128^3$ and $256^3$ boosts to that of the $512^3$ runs, with the grey bands showing the 1\,\% and 2\,\% ranges. The $64^3$ particle simulation is unable to resolve the screening mechanism effectively, due to which the boost for that case rises at intermediate scales. The other two ratios stay within 1\% for $k \lesssim 1 \, h \, \rm{Mpc}^{-1}$ and within 2\,\% for $k<3 \, h \, \rm{Mpc}^{-1}$ and remain of the order of 2\,\% on smaller scales. On linear and quasi-linear scales ($k \lesssim 1 \, h \, \rm{Mpc}^{-1}$), the ratios are very close to unity, displaying the convergence of the boost curves in the linear regime. 

\paragraph{Box size and variance}

\begin{figure}[t]
    \centering
    \includegraphics[width=\hsize]{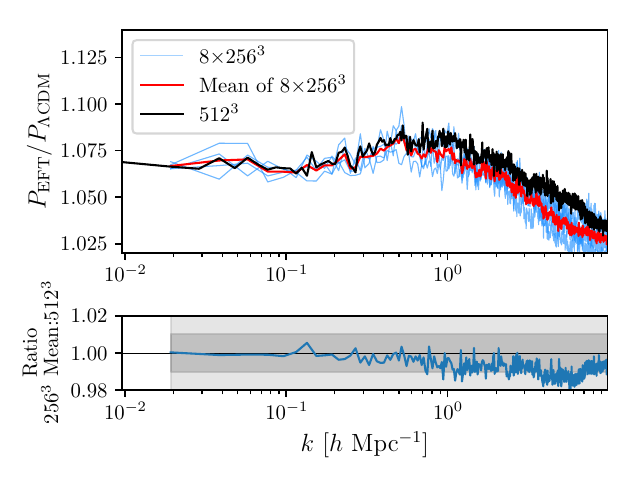}
    \caption{\textbf{Boxsize} \textit{Top}: Boost at $z=0$ for one $512^3$ particle simulation (black), boosts from eight $256^3$ particle simulations (blue), and mean ratio from 8 simulation boxes obtained from splitting the $512^3$ particle simulation (red). \textit{Bottom}: Ratio of the red and black curves from the top panel.}
    \label{fig:conv_boxsize}
\end{figure}

We computed the boost from the $512^3$ particle simulation pair and the eight pairs of $256^3$ particle runs. In the top panel of Fig.~\ref{fig:conv_boxsize}, we show the boost from the $512^3$ version (black) and the mean of the boosts from the $256^3$ simulations (red). We also show the boosts from the eight $256^3$ particle runs in light blue; the spread of these curves displays that the variance is not very large. 

The bottom panel of Fig.~\ref{fig:conv_boxsize} shows the ratio between the red and black curves for range of scales common to the two simulations. The ratio is computed using log-log interpolators for the $512^3$ and $256^3$-particle curves in the top panel. The grey bands show the 1\,\% and 2\,\% levels. The ratio stays within 2\,\% for $k < 9.8\, h \, \rm{Mpc}^{-1}$ and roughly within 1\% for $k\lesssim4 \, h \, \rm{Mpc}^{-1}$, showing the robustness of the results to the size of the simulation box.

\paragraph{Multigrid smoothing cycles}

\begin{figure}[t]
    \centering
    \includegraphics[width=\hsize]{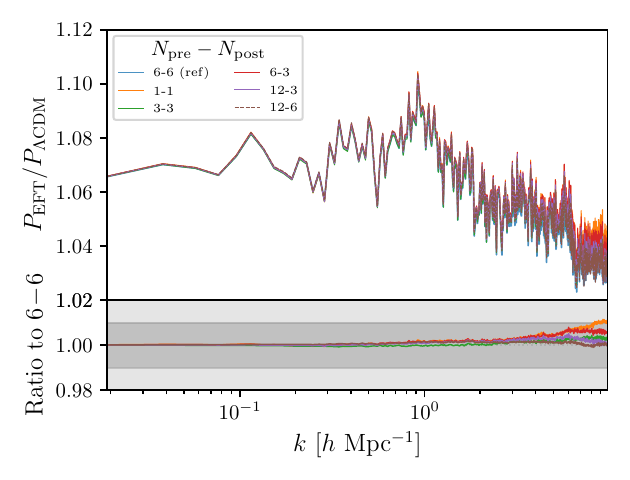}
    \caption{\textbf{Pre- and Post-Smoothing Cycles} \textit{Top}: Power spectrum ratios at $z=0$ for different combinations of $N_{\rm pre}, N_{\rm post}$ for the scalar field multigrid. \textit{Bottom}: The ratio of the curves to the curve for $N_{\rm pre}, N_{\rm post} = 6,6$.}
    \label{fig:conv_nprenpost}
\end{figure}

The top panel of Fig.~\ref{fig:conv_nprenpost} shows the boosts from six paired runs with different $N_{\rm pre}\mbox{--}N_{\rm post}$ combinations. The six curves are extremely close to each other, which is quantified in the bottom panel, which plots the ratios of these curves to the reference curve from the $N_{\rm pre} = N_{\rm post} = 6$ run. The grey bands show the 1\,\% and 2\,\% ranges, as in the other figures. The ratios are all well within 1\% agreement with each other on all scales $k < 10\, h \, \rm{Mpc}^{-1}$ except for 1\mbox{--}1 at small scales. The 3\mbox{--}3, 12\mbox{--}3 and 12\mbox{--}6 curves in the bottom panel are all very close to the reference boost, showing how the solution converges to the 6\mbox{--}6 run at those $N_{\rm pre}\mbox{--}N_{\rm post}$ values. The scalar field solutions have converged well with respect to the number of pre- and post-smoothing cycles. 

\paragraph{Mass refinement threshold}

\begin{figure}[t]
    \centering
    \includegraphics[width=\hsize]{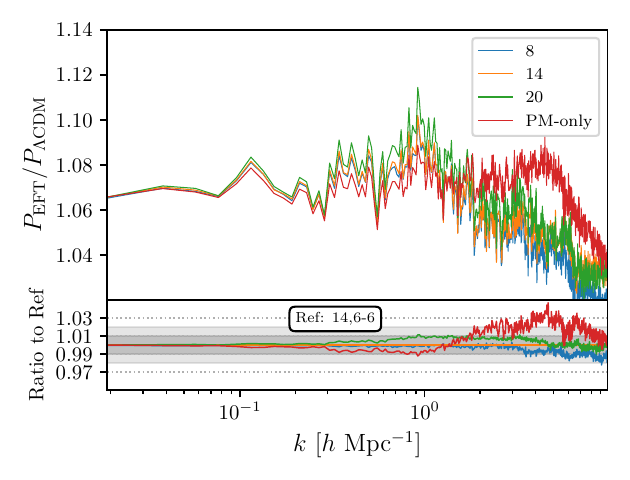}
    \caption{\textbf{Mass Refinement Threshold} \textit{Top}: Power spectrum ratios at $z=0$ with three different $m_{\rm ref}$ values and a PM-only run, with $N_{\rm pre}, N_{\rm post} = 6,6$. \textit{Bottom}: Ratios of the curves to the reference run $m_{\rm ref} = 14$.}
    \label{fig:conv_mref}
\end{figure}

The boosts from the three paired runs ($m_{\rm ref} = \{8,14,20\}$) are plotted in the top panel of Fig.~\ref{fig:conv_mref}, with the colours marking different $m_{\rm ref}$. For comparison, the plot also shows the boost from the particle-mesh run ($m_{\rm ref} = \infty$). The bottom panel shows the ratio of the curves to the reference run with $m_{\rm ref} = 14, N_{\rm pre} = 6, N_{\rm post} = 6$. All the AMR simulation ratios agree to within 1\% of the reference run for all $k \lesssim 5\, h \, \rm{Mpc}^{-1}$ and within 2\% for $k < 9.8\, h \, \rm{Mpc}^{-1}$ . The boost curves for $m_{\rm ref} = 14$ and 8 are very close to each other, showing that the results have converged at $m_{\rm ref} = 14$. Additionally, the boost obtained from the PM-only runs differs from the reference boost by over 3\% for $k \gtrsim 4\, h \, \rm{Mpc}^{-1}$, showing the efficacy of the adaptive mesh refinement at computing the effects of the screening mechanism accurately on intermediate and small scales.

\paragraph{Starting redshift}

\begin{figure}[t]
    \centering
    \includegraphics[width=\hsize]{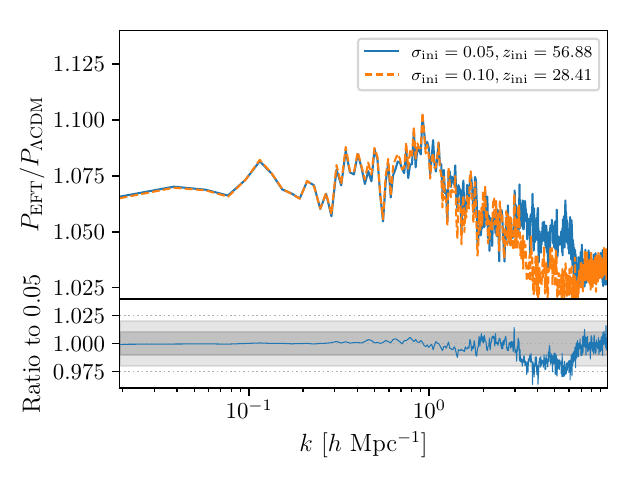}
    \caption{\textbf{Starting redshift} \textit{Top}: Power spectrum ratios at $z=0$ for simulations with different starting redshifts. \textit{Bottom}: Ratio of the two curves.}
    \label{fig:conv_zstart}
\end{figure}

Figure~\ref{fig:conv_zstart} shows the EFT boost curves for the simulation pairs with two starting redshifts in the top panel, with the bottom panel displaying their ratio. The ratio is within 2\,\% for $k \lesssim 4\, h \, \rm{Mpc}^{-1}$ and the two boost curves agree with each other to well within 1\,\% for $k \lesssim 3\, h \, \rm{Mpc}^{-1}$. The ratio remains of the order of 2\,\% or below for $k < 10\, h \, \rm{Mpc}^{-1}$.

\paragraph{Scalar field solver threshold}

\begin{figure}[t]
    \centering
    \includegraphics[width=\hsize]{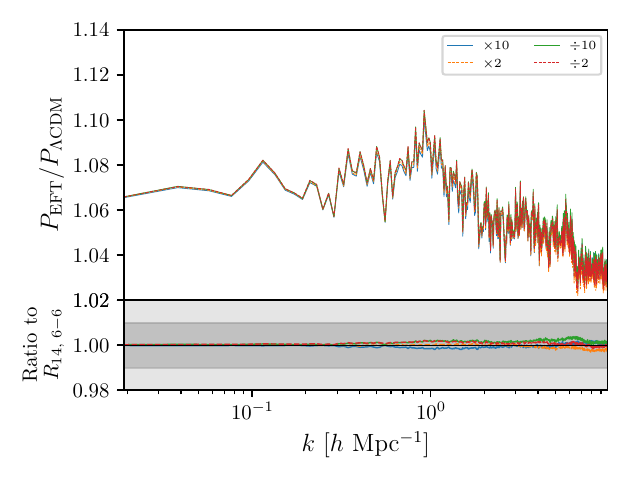}
    \caption{\textbf{Scalar Field Solver Threshold} \textit{Top}: Power spectrum ratios at $z=0$ for runs with the residual error threshold for the scalar field multigrid solver changed by different factors. \textit{Bottom}: Ratios of the curves to baseline threshold simulations.}
    \label{fig:conv_chiconv}
\end{figure}

The top panel of Fig.~\ref{fig:conv_chiconv} shows the boosts from the four simulation pairs with the scalar field solver residual stopping criterion changed by various factors. The bottom panel shows the ratios of these four boost curves to the reference run with the baseline stopping criterion. The ratios agree to well within 1\,\% for all the scales probed by the simulation, demonstrating excellent robustness to the variation of the scalar field convergence threshold. 

Ongoing Stage IV LSS surveys require percent level precision for the analysis of the matter power spectrum. The convergence tests in this section display the robustness of the boost to various numerical parameters, with the boost values mostly staying within 1\% (2\%) of each other for scales with $k \lesssim 3\, h \, \rm{Mpc}^{-1}$ ($k \lesssim 10\, h \, \rm{Mpc}^{-1}$) when the parameters are varied within the converged region (large enough box, small enough mass and spatial resolution etc.). For $\amz = 0$, \citet{Lu:2025gki} constrained $-0.05 \lesssim \abz \lesssim -0.45$ at the 2{--}$\sigma$ level for a $w_0 w_a$CDM cosmology. Our codes provide accuracy to within 1-2\%, which is sufficient in principle to constrain a combination of $\alpha$s of magnitude of the order of $0.1$ to well below the current constraints.

\end{appendix}

\end{document}